\documentclass[11pt]{article}
\usepackage[a4paper,left=0.99in, right=0.99in,top=1.2in, bottom=1.2in]{geometry}

\usepackage[english]{babel}
\usepackage[a4paper]{geometry}
\usepackage{multicol}
\usepackage{graphicx}
\usepackage{amsfonts}
\usepackage{amsmath}
\usepackage{amssymb}
\usepackage{tikz}
\usepackage{subcaption}
\usepackage{xcolor}
\usepackage{hyperref}
\usepackage{cite}

\usetikzlibrary{decorations}
\usetikzlibrary{snakes}

\newcommand{\fww}[0]{\mathcal{F}^{\mathrm{WW}}}
\newcommand{\fdp}[0]{\mathcal{F}^{\mathrm{dipole}}}
\newcommand{\sdp}[0]{\sigma_{\mathrm{dipole}}}
\newcommand{\sdpa}[0]{\sigma_{\mathrm{dipole\,Adj.}}}
\newcommand{\GeV}[0]{\mathrm{GeV}}

\author{
T.~Goda,$^1$, K.~Kutak$^{1}$ and S.~Sapeta$^1$\\\,\\
$^1$ 
{\small\it The H.\ Niewodnicza\'nski Institute of Nuclear Physics PAN,}\\ 
{\small\it Radzikowskiego 152, 31-342 Krak\'ow, Poland}\\
}	 

\title{
Effects of exact kinematics and the Sudakov\\
form factor on the dipole amplitude
}

\date{}

\begin{document}

\maketitle

\vspace{-22em}
\begin{flushright}
  IFJPAN-IV-2023-2
\end{flushright}
\vspace{17.5em}

\begin{abstract}
We investigate effects of exact gluon kinematics on the parameters of the
Golec-Biernat--W\"usthoff, and Bartels--Golec-Biernat--Kowalski saturation
models. The resulting fits show some differences, particularly, in the
normalization of the dipole cross section $\sigma_0$. The refitted models are
used for the dijet production process in DIS to investigate effects of the
Sudakov form factor at Electron Ion Collider energies.  
\end{abstract}

\section{Introduction}

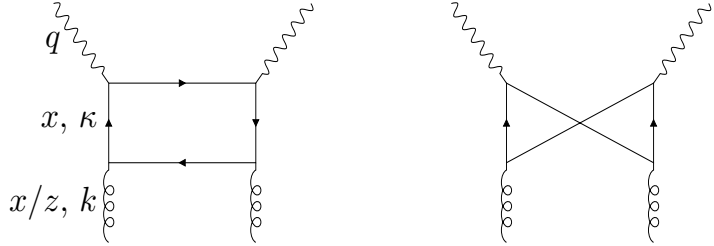
\begin{figure}[t]
\centering
\resizebox{0.6\textwidth}{0.2\textwidth}{
\begin{tikzpicture}
  \draw(-0.75in,0.25in)node[left,scale=0.02in]{$q$};
  \draw[snake=coil,segment aspect=0,segment length=0.1in](-0.875in , 0.5in)--(-0.5in,0.in);
  \draw[snake=coil,segment aspect=0,segment length=0.1in](0.875in ,0.5in)--(0.5in,0.in);
  \draw(-0.5in , 0in)--(0.5in,0in)--(0.5in , -0.5in)--(-0.5in,-0.5in)--cycle ;
  
  \filldraw(0.02in,0)--(-0.02in,-0.02in)--(-0.02in,0.02in)--cycle;
  \filldraw(-0.02in,-0.5in)--(0.02in,-0.52in)--(0.02in, -0.48in)--cycle;
  
  \filldraw(-0.5in,-0.23in)--(-0.52in,-0.27in)--(-0.48in,-0.27in)--cycle;
  \filldraw(0.5in,-0.27in)--(0.52in,-0.23in)--(0.48in,-0.23in)--cycle;
  
  \draw(-0.5in , -0.25in)node[left,scale=0.02in]{$x$, $\kappa$};
  \draw(-0.5in , -0.75in)node[left,scale=0.02in]{$x/z$, $k$};
  \draw[ snake=coil,segment aspect=1.5,segment length=0.1in](0.5in , -1in)--(0.5in,-0.5in);
  \draw[snake=coil,segment aspect=1.5,segment length=0.1in](-0.5in , -1in)--(-0.5in,-0.5in);
  \end{tikzpicture}
  \hspace{60pt}
  \begin{tikzpicture}
  \draw[snake=coil,segment aspect=0,segment length=0.1in](-0.875in , 0.5in)--(-0.5in,0.in);
  \draw[snake=coil,segment aspect=0,segment length=0.1in](0.875in ,0.5in)--(0.5in,0.in);
  \draw(-0.5in , 0in)--(0.5in,-0.5in)--(0.5in ,0in)--(-0.5in,-0.5in)--cycle ;
  
  \filldraw(-0.5in,-0.23in)--(-0.52in,-0.27in)--(-0.48in,-0.27in)--cycle;
  \filldraw(0.5in,-0.23in)--(0.52in,-0.27in)--(0.48in,-0.27in)--cycle;
  
  \draw[ snake=coil,segment aspect=1.5,segment length=0.1in](0.5in , -1in)--(0.5in,-0.5in);
  \draw[snake=coil,segment aspect=1.5,segment length=0.1in](-0.5in , -1in)--(-0.5in,-0.5in);
\end{tikzpicture}
}
\caption{Kinematic variables of the inclusive DIS.}
\label{fig:gluon-kinem}
\end{figure}

Factorization of scales plays central role in {\it Quantum Chromodynamics}
(QCD). In particular, within collinear factorization approach, long-distance
effects can be isolated into objects called {\it collinear parton distribution
functions}~(PDFs)~\cite{Roberts:1990ww,Collins:2011zzd}.  The collinear PDFs are
not fully perturbatively calculable but they obey perturbative evolution
equations and are process-independent~\cite{Collins:2011zzd}. That is to say
that one needs to determine initial condition by fit to data, and the obtained PDFs can be used universally. 

One of the main type of processes that is suited particularly well to the
studies of proton structure is the {\it deep inelastic
scattering}~(DIS)~\cite{Roberts:1990ww,Collins:2011zzd}. The data from HERA has
enabled us to answer many questions in that domain and largely improved our
picture of interior of a proton.  It is therefore very exiting that the next
generation of DIS machines, the Election Ion Collider~(EIC)~\cite{NAP25171} is
making its way and will soon allow us to uncover more details about hadron
structure.

At large center of mass energy and fixed value of the photon virtuality, $Q$,
one probes  the region of small Bjorken scaling variable, $x$.  In that region,
which is dominated by gluons~\cite{Roberts:1990ww,Collins:2011zzd,
Kovchegov:2012mbw}, the approach of High Energy Factorization, also called
$k_T$-factorization~\cite{Catani:1990xk,Catani:1990eg,Collins:1991ty,Catani:1993ww,Catani:1993rn,Catani:1994sq},
has proven to be suited particularly well.  In this framework, the interaction
of the photon with gluons happens through  boson-gluon fusion or quark impact factor, as depicted in
Fig.~\ref{fig:gluon-kinem}. An object, which is central to the description of
such process is called the dipole gluon density, $\fdp(x,k_T^2)$, and it is a
type of {\it transverse momentum
dependent}~(TMD)~\cite{Bomhof:2006dp,Dominguez:2010xd,Dominguez:2011wm,Xiao:2017yya}
PDF. 
Due to its clear picture, one often uses a position-space counterpart of the
gluon density, called {\it dipole cross section}. 
It appears in the $k_T$-factorization
formula~\cite{Catani:1990eg,Dominguez:2010xd,Dominguez:2011wm,Xiao:2017yya},
which we shall discuss in the next section, and plays a role similar to that of
the collinear parton distribution function in the collinear factorization.  

Much effort has been made to study this object, notably the work of Balitsky,
Fadin, Kuraev and Lipatov (BFKL)~\cite{Balitsky:1978ic, Kuraev:1977fs} has
predicted sharp rise of cross sections with $1/x$.  However, such rise violates
unitarity and the Froissart
bound~\cite{Roberts:1990ww,Collins:2011zzd,Kovchegov:2012mbw,Barone:1993sy}. In
Ref.~\cite{Gribov:1983ivg} it has been recognized that gluon recombination can
tame the growth of gluons. The interplay of linear and nonlinear terms leads to
a phenomenon called {\it gluon saturation} and corresponding evolution equations
are known as Balitsky-Kovchegov~(BK)~\cite{Balitsky:1995ub,Kovchegov:1999yj}
equation or
JIMWLK\cite{Kovner:1999bj,Kovner:2000pt,Iancu:2000hn,JalilianMarian:1997dw,JalilianMarian:1997gr,JalilianMarian:1997jx}
equation.

Despite the success of the aforementioned evolution equations, there exist a
number of phenomenological models of the dipole cross section, which are popular
due to to their simplicity.  A notable example is the model proposed by
Golec-Biernat and Wu\"sthoff (GBW)~\cite{Golec-Biernat:1998zce} and its
extension by Bartels, Golec-Biernat and Kowalski (BGK)~\cite{Bartels:2002cj}. We
will use these two specific models in our study. We note, however, that other
saturation models have been discussed in the
literature~\cite{Kowalski:2003hm,McLerran:1993ni,Forshaw:1999uf,Iancu:2003ge}.

In the present study, we fit the GBW and BGK models to HERA
data~\cite{Abt:2017nkc} in the $k_T$-factorization formula for $F_2$, instead of
the original dipole factorization~\cite{Mueller:1994jq} formula. As discussed below, this allows us to
relatively easy investigate the role of the  exact gluon kinematics.  Since
inclusive observables do not reveal the $k_T$ dependence well, in order to
demonstrate the effects of exact kinematics we consider a less inclusive
observable which is more sensitive to the $k_T$ dependence of the gluon
distribution.  The resulting models are used for predictions of dijet production
in DIS at Electron Ion Collider (EIC) \cite{AbdulKhalek:2021gbh}, including effects of the Sudakov form factor, which resums logarithms of the small transverse momentum. 
This process is known to be sensitive to another type of gluon density, called
Weitzs\"acker-Williams~(WW)~\cite{Dominguez:2010xd,Dominguez:2011wm}. For that
process, we study, in particular, the angular correlation of jets, and that of
the scattered electron and the jets. For other papers addressing dijet production at the EIC we refer the Reader to~\cite{Caucal:2023nci,Caucal:2021ent,Boussarie:2021ybe, Taels:2022tza,Altinoluk:2018byz}. 
 
The paper is organized as follows.  In the next section, we present theoretical
framework which we use and outline differences between the $k_T$-factorization
and the dipole factorization.  In Sec.~\ref{sec:F2fits}, results of the new fits
to HERA data in the $k_T$-factorization formula are presented.  In
Sec.~\ref{sec:dijetsEIC}, we apply the results from the previous section to
compute distributions for dijet production in DIS at the EIC and make
comparisons to our earlier study of Ref.~\cite{vanHameren:2021sqc}.

\section{The framework}

The DIS cross section (structure function) factorises and can be written in a
form
\begin{equation}
   d\sigma = \sum_{a}  \phi_{a/h} \otimes H_{\gamma a \to X} ,
\end{equation}
where the hard function, $H_{\gamma a \to X}$, involving a parton $a$ in the
initial state, is calculable perturbatively, and $\phi_{a/h}$ is the parton
distribution function of $a$ in a hadron $h$.  
The symbol $\otimes$ denotes appropriate convolution. All nonperturbative
effects are absorbed in $\phi_{a/h}$, while $H_{\gamma a \to X}$ can be computed
order by order in $\alpha_s$.  
 
We will study factorization in two versions: defined in momentum and position
space, respectively.  As mentioned earlier, in the fit, we use the GBW and BGK
models.  The models were originally formulated in the position-space version of
the $k_T$-factorization formula.  In the GBW model, the dipole cross section has
the form~\cite{Golec-Biernat:1998zce}
\begin{align}
\label{eq:gbw-dipole}
\sigma_{\mathrm{GBW}}(x,r)&=\sigma_0\left(1-e^{-r^2/R^2_0}\right)&\mathrm{where}&
&R^{-2}_0&=\frac{Q_0^{2}}{4}\left(\frac{x_0}{x}\right)^\lambda.
\end{align}
The dipole cross section is related to the dipole gluon density by the Fourier
transform
\begin{equation}
\alpha_s\fdp(x,k_T^2)=\frac{N_c}{4\pi}\int\frac{d^2\mathbf{r}}{(2\pi)^2}e^{i\mathbf{k}_t\cdot \mathbf{r} }\nabla_{\mathbf{r}}^2\sigma_{\mathrm{dipole}}(x,r).
\label{eq:dipole gluon}
\end{equation}
The essence of the GBW model is encoded in the $x$-dependent saturation scale
$Q_s^2(x)=Q^2_0(x_0/x)^{-\lambda}$, which separates the saturation region and
the scaling region.  An extension of the above model was proposed by Bartels,
Golec-Biernat and Kowalski~\cite{Bartels:2002cj} who incorporated the DGLAP
evolution in the GBW dipole cross section~(\ref{eq:gbw-dipole}) by modifying the
exponent to 
\begin{equation}
R_0^{-2}=\frac{\pi^2\alpha(\mu^2)xg(x,\mu^2)}{3\sigma_0},
\end{equation}
where
\begin{align}
\mu^2&=\frac{C}{r^2}+\mu_0^2 & &\mathrm{and} & xg(x,Q^2_0)&=A_g x^{-\lambda_g}(1-x)^{5.6},
\end{align}
thus improving the description of data at higher $Q^2$.

While these models enjoyed much success in the phenomenology of DIS, including
the diffractive and photo-production
processes~\cite{Golec-Biernat:1998zce,Golec-Biernat:1999qor}, it is worth
mentioning that they both use certain kinematic approximation which is specific
to LO dipole factorization  and are not there in the $k_T$-factorization. This
was recognized in~\cite{Bialas:2000xs}. In the following, we will investigate
effects of these approximations.

Firstly, let us outline the factorization formulas.  With $q'\equiv q+x p$, one
decomposes $k$ and $\kappa$, defined in Fig.~\ref{fig:gluon-kinem}, as
\begin{align}
    \kappa&=\alpha p-\beta q'+\kappa_t&&\mathrm{and}& k&=a p- bq'+k_T.
\end{align}
For the contribution depicted in Fig.~\ref{fig:gluon-kinem}, the structure
function $F_2$ factorizes with a change of variable
${\boldsymbol{\kappa}'}_t\equiv{\boldsymbol{\kappa}_t}-(1-\beta)\mathbf{k}_t$,
to the form~\cite{ Kimber:2001uaa,Kwiecinski:1997ee}
\begin{multline}
	F_2(x,Q^2)=\sum_f e_f^2 \frac{Q^2}{2\pi}\int\frac{dk^2_t}{k_T^2}\int^1_0d\beta\int d{\kappa'}_t\alpha_s(\mu^2) \fdp(x/z,k_T^2)\Theta(1-x/z)\\
	\left[\left(\beta^2+(1-\beta)^2\right)\left(\frac{I_1}{2\pi}-\frac{I_2}{2\pi}\right)
	+\left(m_f^2+4Q^2\beta^2(1-\beta)^2\right)\left(\frac{I_3}{2\pi}-\frac{I_4}{2\pi}\right)\right],
	\label{eq:angle-integrated}
\end{multline}
where
\begin{equation}
	\frac{1}{z}=1+\frac{{\kappa'}^2_t+m_f^2}{\beta(1-\beta)Q^2}+\frac{k^2_t}{Q^2}\,.
	\label{eq:z}
\end{equation}

One should note that the gluons are not probed directly, thus the argument of
the gluon density is $x/z$ rather than $x$.  If one, instead, uses $ 1/z=1+4
m_f^2/Q^2$ and assumes that $\mu$ is independent of $\kappa'$ and $k_T$, the
above formula can be written in the impact parameter
space~\cite{Golec-Biernat:1998zce,Nikolaev:1990ja} ({\it i.e.} as a dipole
factorization formula)
\begin{equation}
  F_2\left(x,Q^2\right)=\sum_f e_f^2 \frac{Q^2}{4\pi^2} \int^1_0d\beta \int \frac{d^2\mathbf{r}}{(2\pi)^2} \left|\Psi\left(\tilde{x},\beta,Q^2\right)\right|^2\sigma_{\mathrm{dipole}}\left(\tilde{x},r\right),
  \label{eq:dipole factorization}
\end{equation}
where the photon wave function, $\left|\Psi\left(x,\beta,Q^2\right)\right|^2$,
describes splitting of the incoming photon into a $q\overline{q}$ pair with
light-cone momenta fractions $\beta$ and $1-\beta$ respectively, and the dipole
cross section, $\sigma_{\mathrm{dipole}}\left(\tilde x,r\right)$, describes the
interaction of the colour dipole of size $r$ with the proton. In
Ref.~\cite{Golec-Biernat:1998zce}, $\tilde{x}=x(1+4m_f/Q^2)$ with nonzero value
of $m_f$ was used even for light flavours, which is necessary to partially
substitute $k_T$ and ${\kappa'}_T$ in Eq.~(\ref{eq:z}).  In the present study,
we use Eq.~(\ref{eq:angle-integrated}) to fit the models, where the gluon
density is obtained by evaluating Eq.~(\ref{eq:dipole gluon}).

Considering the small-$r$ limit of the GBW model
\begin{equation}
    \left.\sigma_{\mathrm{GBW}}\left(x,r\right)\right|_{r\ll Q_s}\approx r^2
    Q_s^2/4\,,
\end{equation}
comparison to 
\begin{equation}
\left. \sigma_{\mathrm{dipole}}\left(x,r\right)\right|_{r\ll
Q_s}\approx\frac{r^2 \pi^2\alpha_s(\mu^2)xg(x,\mu^2)}{3}\,,
\end{equation} 
from BGK, suggests that the GBW model is an approximation in which
$\alpha_s(\mu^2)$ and $xg(x,\mu^2)$ are independent of
$r$~\cite{Bartels:2002cj}.  
For this reason, in one version of the model discussed below, we shall account
for the running coupling in Eq.~(\ref{eq:angle-integrated}) by assuming that
$\alpha_s$ in Eq.~(\ref{eq:dipole gluon}) is constant for the GBW model, and
thus explicitly multiply it by the running coupling
\begin{equation}
\alpha_s(\mu^2)\fdp(x,k_T^2)\rightarrow \alpha_s(\mu)\frac{\alpha_s\fdp(x,k_T^2)}{0.2},
\end{equation}
where
\begin{equation}
\alpha_s(\mu^2)=\frac{1}{\frac{11 C_A-2n_f}{12\pi}\log\left(\frac{\mu^2}{\Lambda_{\mathrm{QCD}}^2 }\right)},
\end{equation}
and  $\Lambda_{\mathrm{QCD}}^2=0.09\;\GeV^2$.

The factor 0.2 is an arbitrary normalization, whose effect is absorbed by
$\sigma_0$, and hence bears no importance.  (For a more detailed analysis of the
dipole gluon density from the BGK model see
Ref.~\cite{Luszczak:2022fkf}\footnote{In Ref.~\cite{Luszczak:2022fkf}, the
coupling constant was treated differently and the large $k_T$-region was
matched to the derivative of $xg(x)$, while we have used series transformation
to accelerate the high-$k_T$ integration. Their treatment is also different from
the original BGK paper~\cite{Bartels:2002cj}. As a consequence, their gluon
density is positive in the large $k_T$-region and the overall large-$k_T$
behaviour is significantly different.}) While there is some ambiguity on what
the argument of $\alpha_s(\mu^2)$ should be, we follow
Ref.~\cite{Kwiecinski:1997ee} and use
\begin{equation}
	\mu^2=k_T^2+{\kappa'}_t^2+m_f^2+\mu^2_0,
\end{equation}
where we add $\mu^2_0=4\;\GeV^2$ in order to freeze the coupling at low scales.

\section{Fits to $F_2$ data}
\label{sec:F2fits}

We fitted the GBW and BGK models to the $F_2$ data from
HERA~\cite{Abt:2017nkc}\footnote{For description of $F_2$ at NLO accuracy within
dipole factorisation see Ref.~\cite{Beuf:2020dxl} }.
A numerical program was written with help of
CERNLIB~(DPSIPG)~\cite{Kolbig:1972zz}, GSL~\cite{GSL}, CUBA~\cite{Hahn:2004fe}
and ROOT~\cite{Brun:1997pa} libraries to evaluate $F_2$. The fitting was
performed using MnMigrad and MnSimplex of ROOT::Minuit2~\cite{James:2004xla}.

The data were selected to be in the range $0.045\leq Q^2\leq
650\;\mathrm{GeV^2}$, $x<0.01$.  As with the previous fit~\cite{Goda:2022wsc},
the $c$ and $b$ flavours were taken into account with the mass $1.3$ and
$4.6\;\mathrm{GeV}$, respectively. As discussed earlier, we take the light
quarks to be massless.  

We studied the following cases
\begin{itemize}
  \item 
  GBW model with the fixed coupling in $k_T$-factorization  ($k_T$-GBW),
  \item 
  GBW model with the running coupling in $k_T$-factorization (rc-$k_T$-GBW),
  \item 
  BGK model in $k_T$-factorization ($k_T$-BGK),
\end{itemize} 
and, as a reference, we provide the following results from
Ref.~\cite{Goda:2022wsc}
\begin{itemize}
  \item
  GBW model with massless light quarks in the dipole factorization ($r$-GBW),
  \item
  GBW model with massive light quarks in the dipole factorization
  ($r$-GBW-massive),
  \item
  BGK model in  the dipole factorization ($r$-BGK).
\end{itemize} 

\begin{table}[t]
\begin{subtable}{\textwidth}
\center\footnotesize
\begin{tabular}{|c|c|c|c|c|}
	\hline - & $\sigma_0 $ [mb] & $x_0 \left(10^{-4}\right)$ & $\lambda$ & $\chi^2/\mathrm{dof}$ \\\hline 
	{\footnotesize $r$-GBW} & 1.907e+01& 2.582e+00& 3.219e-01& 4.438e+00\\\hline 
	{\footnotesize $r$-GBW-massive} & 2.384e+01& 1.117e+00& 3.082e-01& 5.274e+00\\\hline 
	{\footnotesize $k_t$-GBW} & 3.344e+01& 1.333e+00& 3.258e-01& 4.396e+00\\\hline 
	{\footnotesize rc-$k_t$-GBW} & 1.520e+01& 2.648e+00& 3.211e-01& 2.447e+00\\\hline 
\end{tabular}
\vspace{2mm}
\end{subtable}
\begin{subtable}{\textwidth}
\center\footnotesize
\begin{tabular}{|c|c|c|c|c|c|c|}
	\hline - & $\sigma_0 $ [mb] & $A_g$ & $\lambda_g$ & $C$ & $\mu_0^2 \left[\mathrm{GeV^2}\right]$ & $\chi^2/\mathrm{dof}$ \\\hline 
	{\footnotesize $r$-BGK} & 2.326e+01& 1.181e+00& 8.317e-02& 3.294e-01& 1.873e+00& 1.556e+00\\\hline 
	{\footnotesize $k_t$-BGK} & 3.470e+01& 1.048e+00& 2.205e-01&2.391e-01& 9.954e-01&  1.527e+00\\\hline 
\end{tabular}
\vspace{2mm}
\end{subtable}
\caption{Fit parameters of respective models. The parameters of the
dipole factorization cases are from Ref.~\cite{Goda:2022wsc}.}
\label{tab:table}
\end{table}

The results of the fits are summarized in Tab.~\ref{tab:table}.  The fit quality
of $k_T$-GBW is almost unchanged w.r.t. $r$-GBW, while rc-$k_T$-GBW shows
remarkable improvement, almost halving the $\chi^2$ value.  This is in line with
the observation made in
Refs.~\cite{Albacete:2004gw,Albacete:2007yr,Albacete:2010sy} that in the BK
evolution, the running coupling corrections have considerable effect.

Another notable point is that, except for the normalization $\sigma_0$, the
parameters are very similar, particularly those of rc-$k_T$-GBW are almost
identical to those of $r$-GBW. While the GBW model remains almost unaffected,
the BGK model seems to show slightly more change. The difference in $\sigma_0$
is similar to that of GBW and other parameters changed moderately.

\begin{figure}[p]

  \begin{multicols}{2}
  \begin{subfigure}{0.5\textwidth}
    \includegraphics[width=\textwidth]{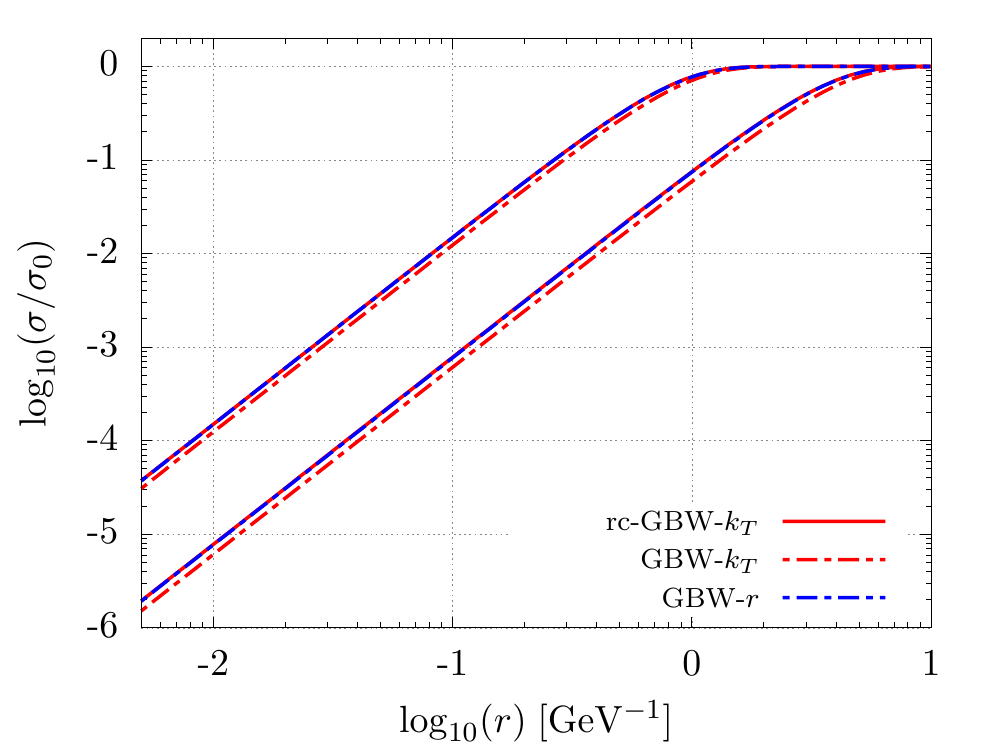}
    \includegraphics[width=\textwidth]{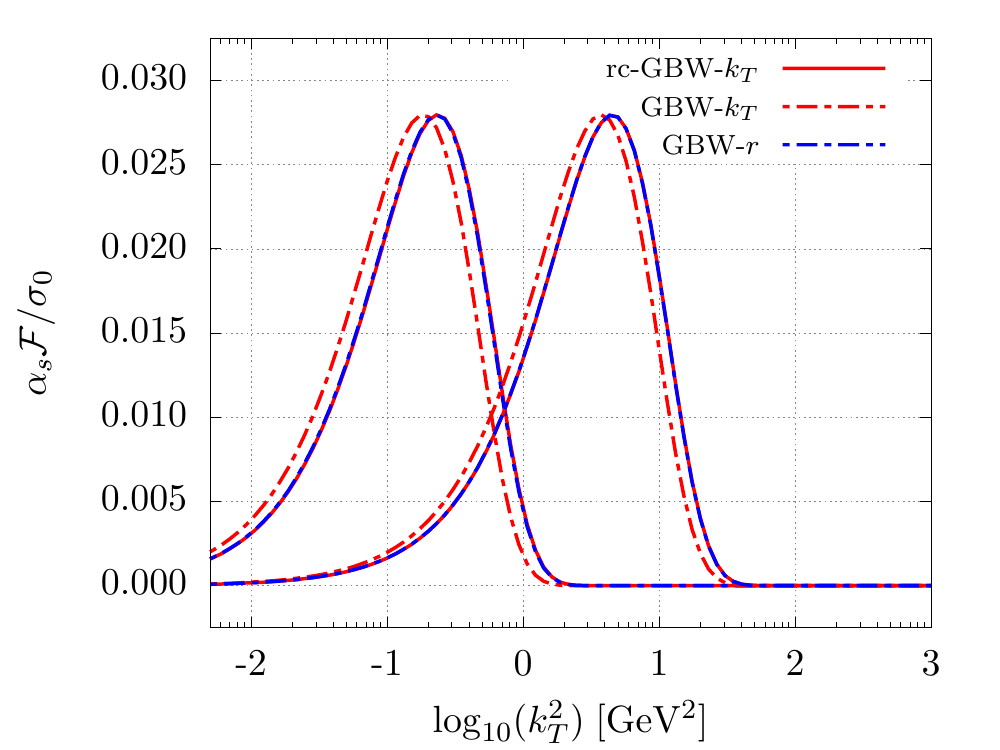}
    \includegraphics[width=\textwidth]{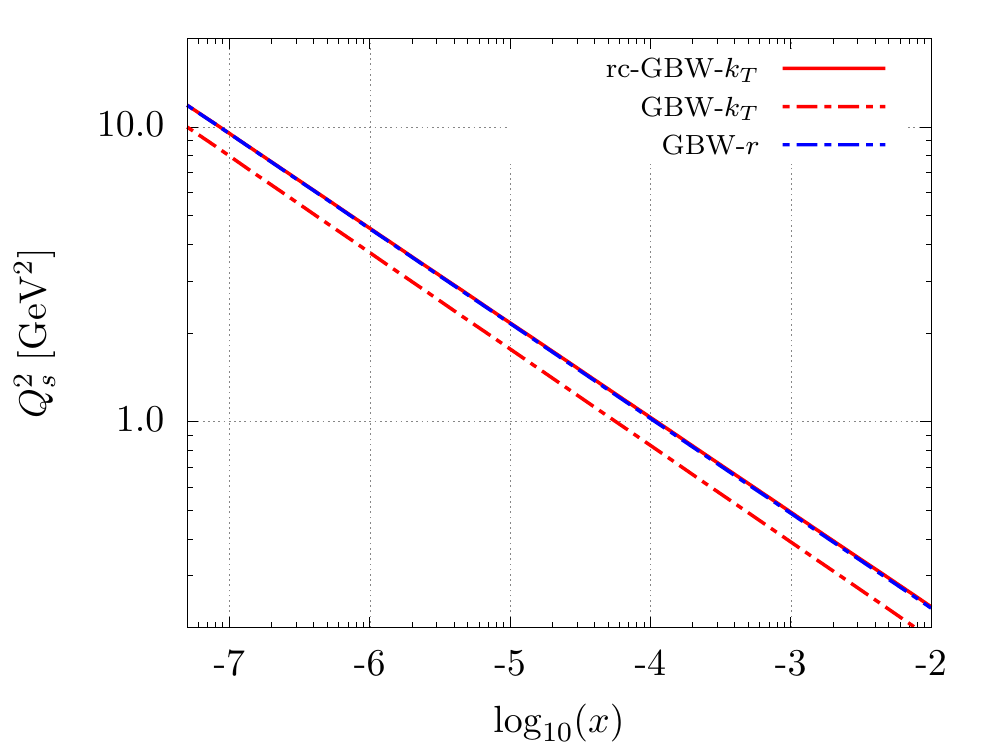}
    \caption{GBW}
    \label{fig:GBW}
\end{subfigure}
\begin{subfigure}{0.5\textwidth}
    \includegraphics[width=\textwidth]{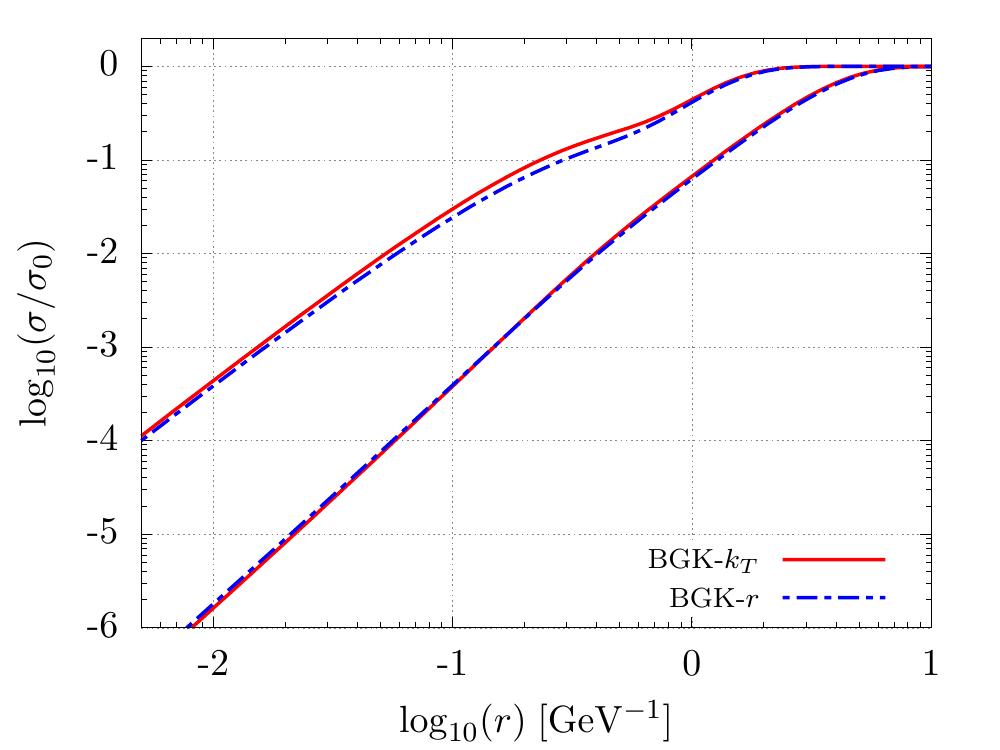}
    \includegraphics[width=\textwidth]{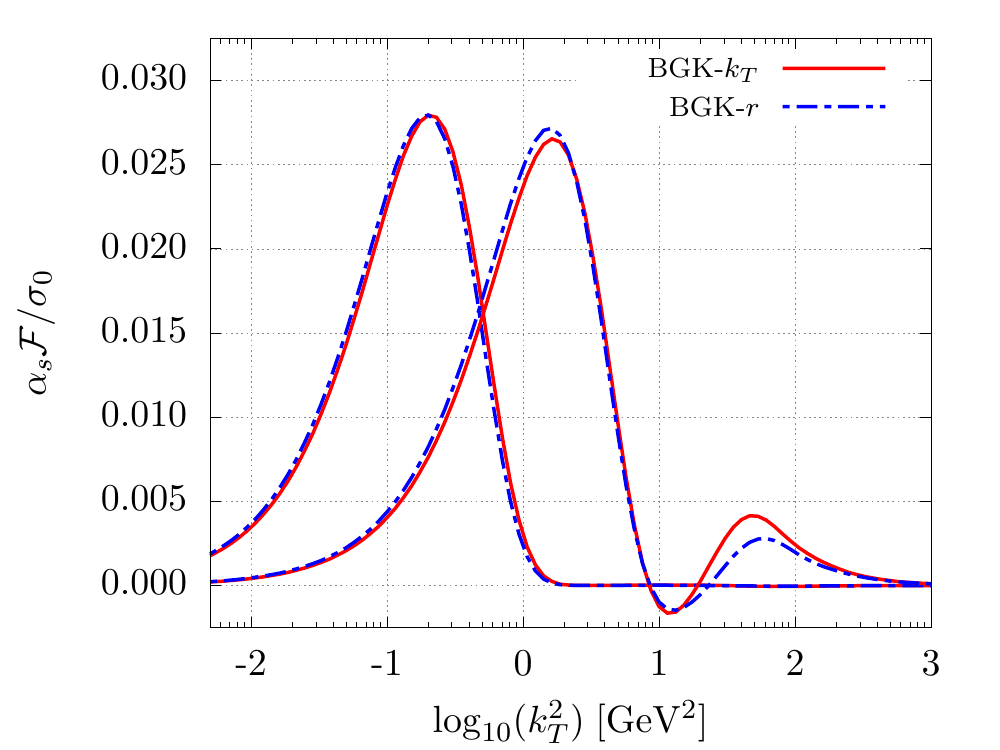}
    \includegraphics[width=\textwidth]{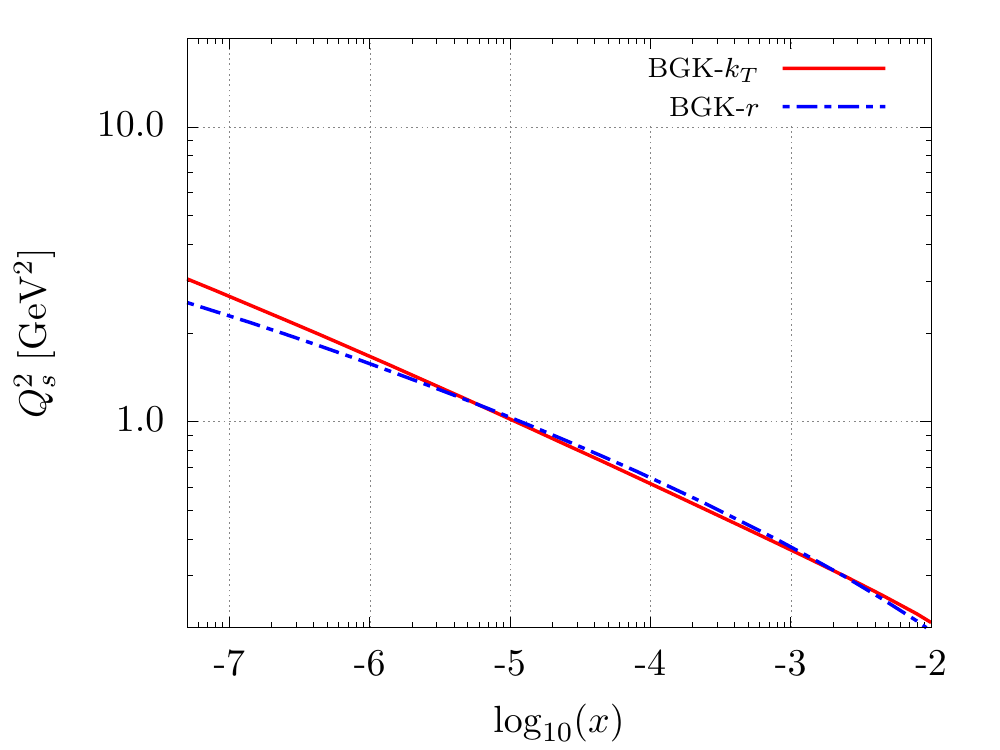}
    \caption{BGK}
    \label{fig:BGK}
\end{subfigure}
  \end{multicols}
  \caption{The dipole cross section, the dipole gluon density at $x=10^{-2},\;
  10^{-6}$, and the saturation scale for the GBW (left) and the BGK (right)
  models. Note that the dipole cross section and the gluon density are
  normalized with $\sigma_0^{-1}$.} 
\label{fig:GBW-BGK}
\end{figure}

In Fig.~\ref{fig:GBW-BGK}, the plots of the dipole cross section, the dipole
gluon density and the saturation scale for the GBW and BGK models are shown. The
dipole cross section and the gluon density are both normalized by $\sigma_0$ in
order to show the effects  of other parameters better, and the saturation scale
is defined as a ridge of the dipole gluon density in the $(x,k_T^2)$ plane.

In the plots on the left hand side, one can see the effects of changes in the
parameters of the GBW model. The difference between the rc-$k_T$-GBW and $r$-GBW
is negligible, while the $k_T$-GBW is slightly shifted, compared to others, by the change in $x_0$.

On th right hand side of Fig.~\ref{fig:GBW-BGK}, the same plots are shown for
the BGK model. Unlike in the GBW case, the connection between the differences
shown in the plots and the differences in the parameters is less clear.
Nevertheless, the differences are more prominent in the small-$x$ region.

\begin{figure}[p]
  \includegraphics[width=0.49\textwidth]{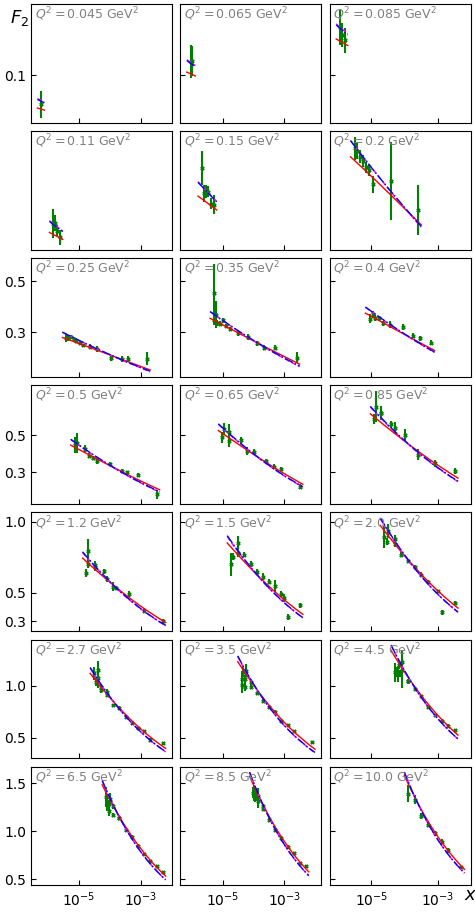}
  \includegraphics[width=0.49\textwidth]{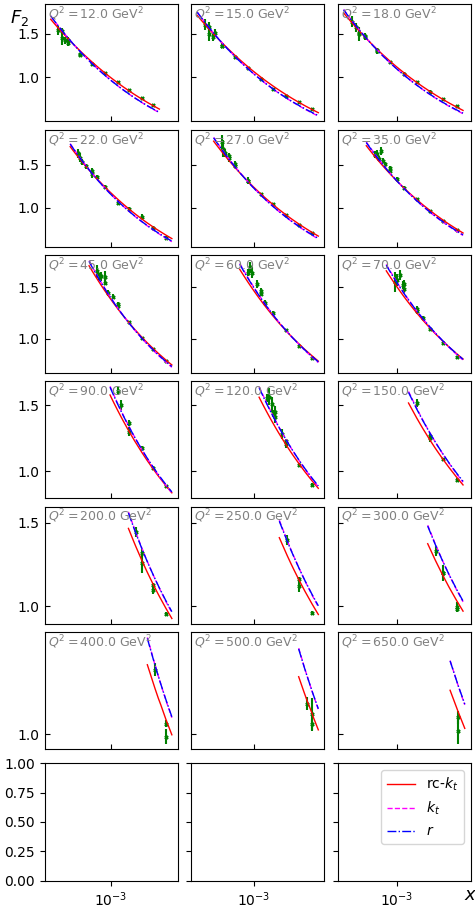}
  \includegraphics[width=\textwidth]{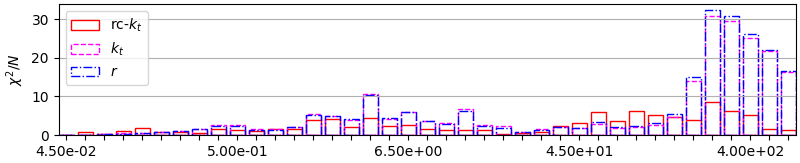}
  \caption{Comparison of $F_2$ fro GBW with HERA data. The histogram shows the
  $\chi^2$ value per data point in each frame.  Improvement by the running
  coupling (rc-$k_T$) is clearly visible in the high-$Q^2$region, while the new
  fit ($k_T$) shows only marginal improvement.}
  \label{fig:GBW-Grid}
\end{figure}

\begin{figure}[p]
  \includegraphics[width=0.49\textwidth]{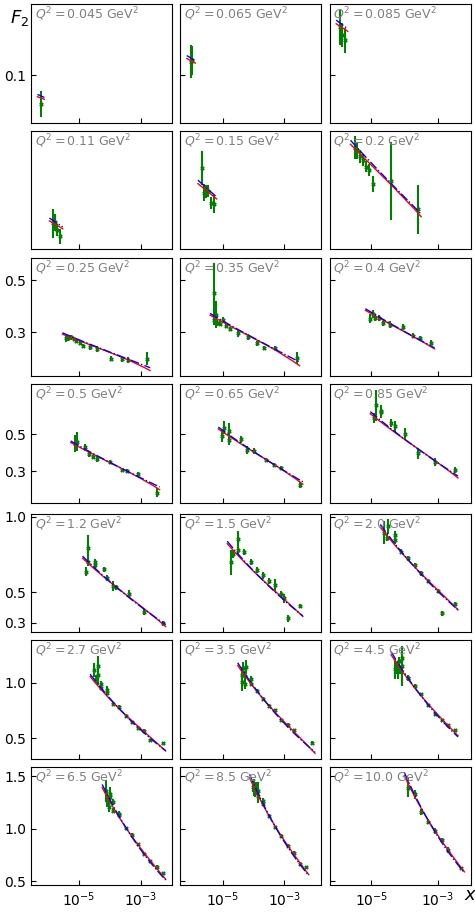}
  \includegraphics[width=0.49\textwidth]{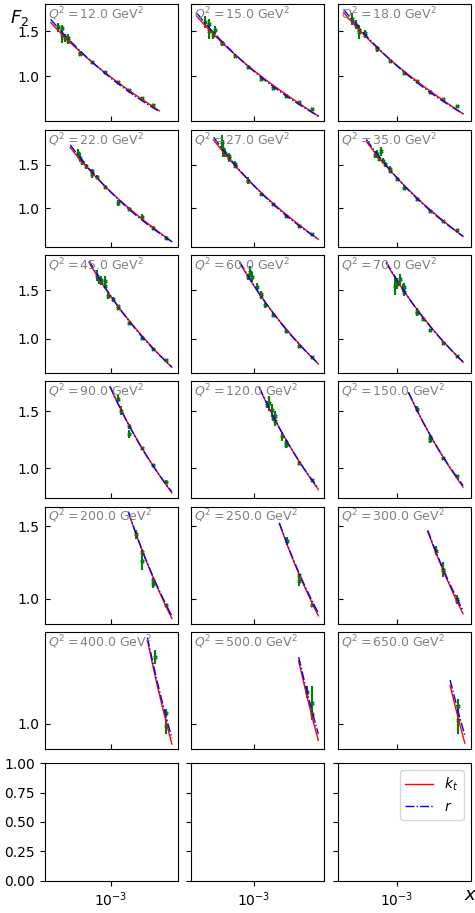}
  \includegraphics[width=\textwidth]{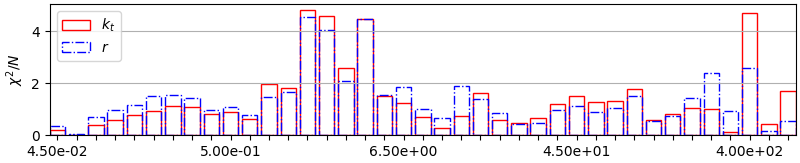}
  \caption{Comparison of $F_2$ from BGK with HERA data. The histogram shows the
  $\chi^2$ value per data point in each frame. The overall fit quality remain
  similar but the quality in each frame changes noticeably. In particularly, the
  $k_T$-formula ($k_T$) shows better quality at small $Q^2$.}
  \label{fig:BGK-Grid}
\end{figure}

Comparisons of the results with the $F_2$ data are shown in
Figs.~\ref{fig:GBW-Grid} and~\ref{fig:BGK-Grid}.  In Fig.~\ref{fig:GBW-Grid},
the differences between $r$-GBW and $k_T$-GBW are not visible, while
rc-$k_T$-GBW shows sizable difference from the others, particularly in the
large-$Q^2$ region.  Recalling that the parameters of rc-$k_T$-GBW and $r$-GBW
are very similar, the difference in $F_2$ is almost entirely due to the coupling
constant. The improvement in the fit quality is depicted as a histogram of the
$\chi^2$-vale per number of points at the bottom of Fig.~\ref{fig:GBW-Grid}.
Here, the improvement in the large-$Q^2$ region is very clear.  In
Fig.~\ref{fig:BGK-Grid}, the differences between $r$-BGK and $k_T$-BGK are
hardly visible. In the histogram at the bottom, one can see some differences,
but they cancel out mostly, making little improvement overall.

Let us now take a closer look at the parameter $\sigma_0$. Recall that the
difference between the $k_T$-factorization formula and the dipole factorization
formula is in $x/z$, and this enters in the GBW formalism as $\tilde{x}$. It is
easy to see that, as $x$ grows, the dipole cross section gets suppressed.
(Keeping in mind the suppression by the photon wave function in the large-$r$
region.) Such effect was discussed previously in Ref.~\cite{Kutak:2004ym} in the
context of the BK equation. In fact, this suppression is the motivation given in
Ref.~\cite{Golec-Biernat:1998zce} for such modification of $x$, so that, in the
small-$Q^2$ limit, the total cross section remains finite. Since 
\begin{equation}
\left(1+\frac{4 m_f^2}{Q^2}\right)\leq\left(1+\frac{k_T^2}{Q^2}+\frac{{\kappa'}_t^2+m_f^2}{\beta(1-\beta)Q^2}\right),
\end{equation}
the $k_T$-factorization case receives more suppression. Consequently, the
normalization factor $\sigma_0$ rises to compensate the suppression.  Therefore,
one can understand the change in $\sigma_0$ as a direct consequence of the key
difference between the $k_T$-factorization and the dipole factorization.

\section{Dijet production at EIC}
\label{sec:dijetsEIC}

\begin{figure}[t]
    \begin{subfigure}{0.5\textwidth}
        \includegraphics[width=\textwidth]{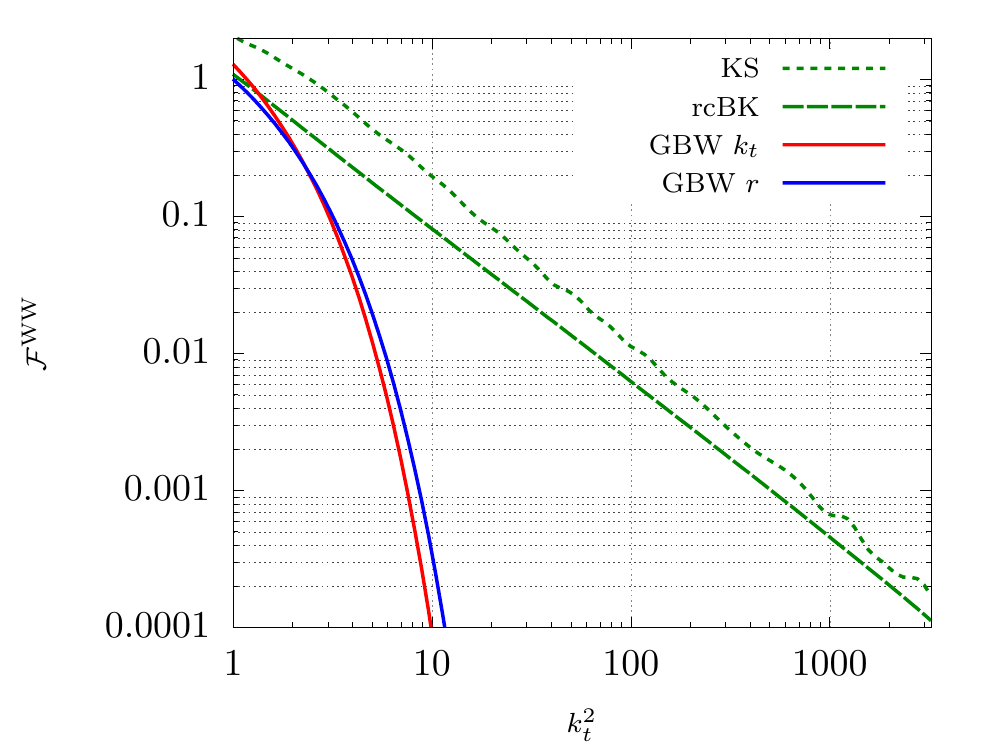} 
    \end{subfigure}
    \begin{subfigure}{0.5\textwidth}
        \includegraphics[width=\textwidth]{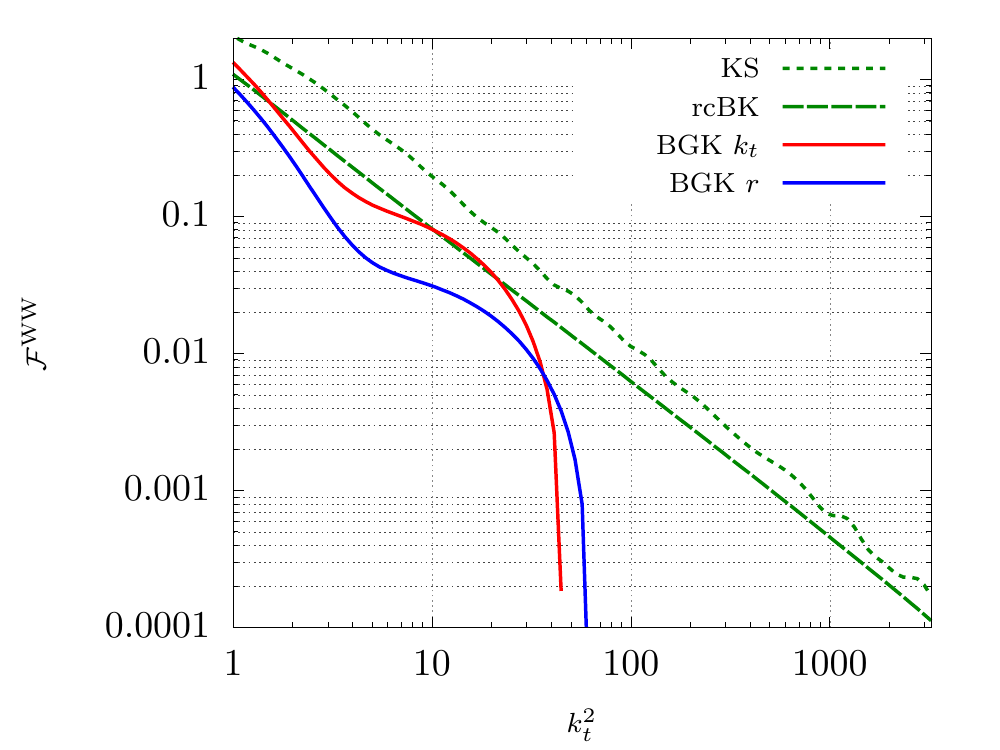} 
    \end{subfigure}
    \begin{subfigure}{0.5\textwidth}
        \includegraphics[width=\textwidth]{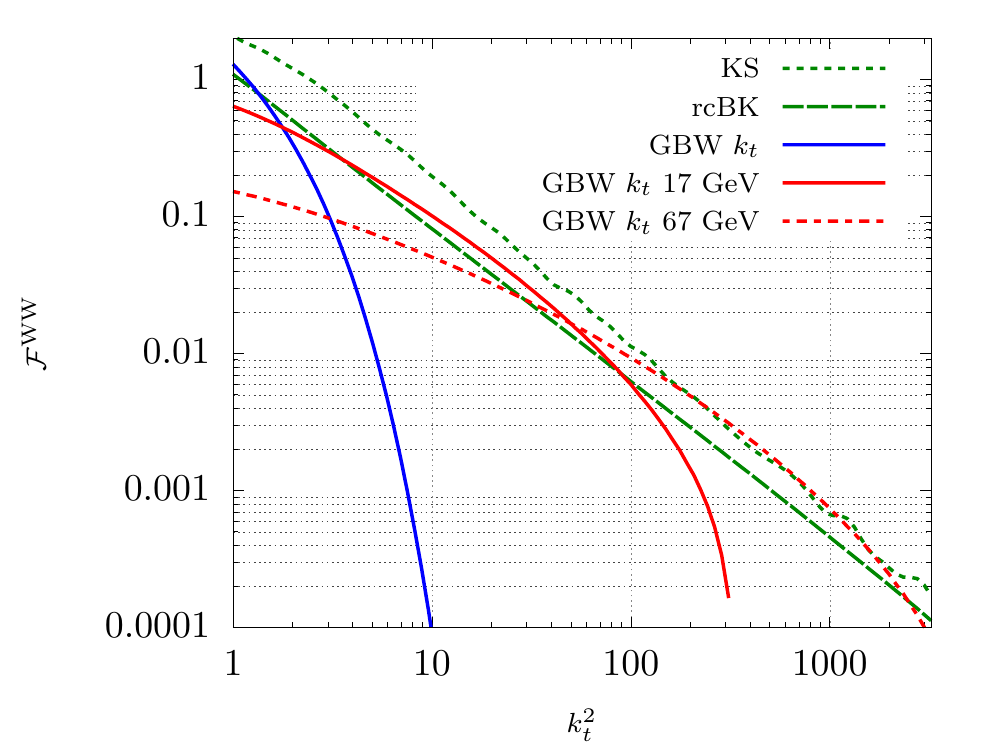} 
    \end{subfigure}
    \begin{subfigure}{0.5\textwidth}
        \includegraphics[width=\textwidth]{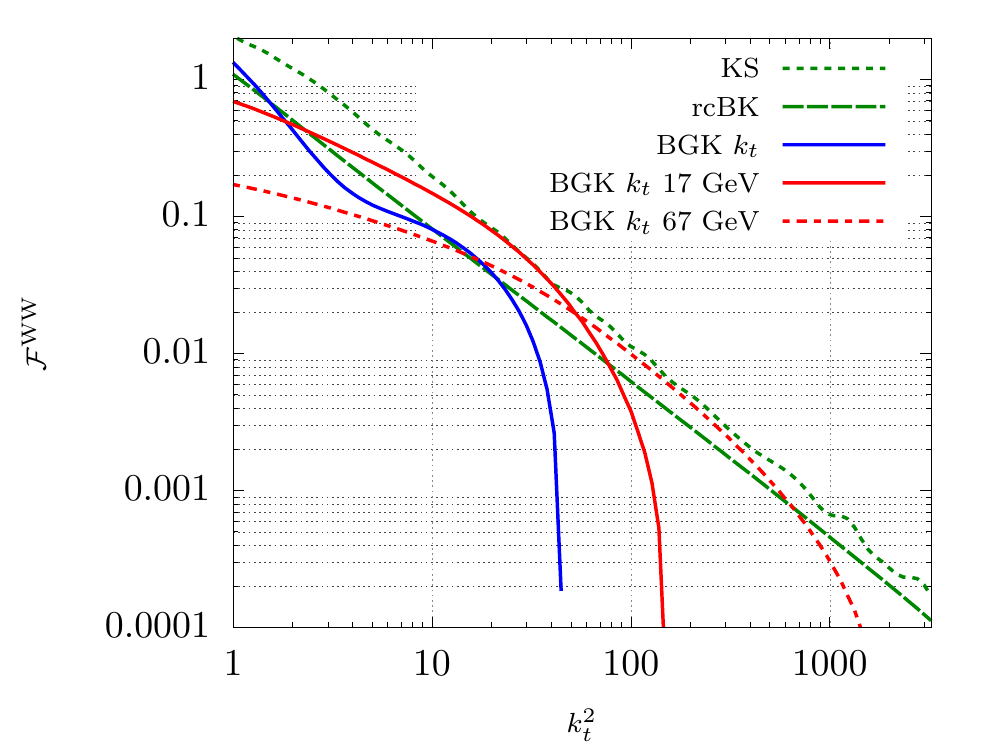} 
    \end{subfigure}
    \caption{ Weizs\"acker-Williams gluon density at $x=10^{-3}$. Top row:
    comparison of the dipole factorization fit and $k_T$-factorization fit
    results. Bottom row: comparison of the respective models with and without
    the Sudakov factor, at $\mu=17,\;67\ \mathrm{GeV}$.The green dotted line is
    the KS gluon~\cite{vanHameren:2021sqc}, and the green dashed line is the
    rcBK gluon~\cite{Hentschinski:2022rsa}.} 
    \label{fig:ww}
\end{figure}

\begin{figure}[p]
	\begin{subfigure}{0.5\textwidth}
	\includegraphics[width=\textwidth]{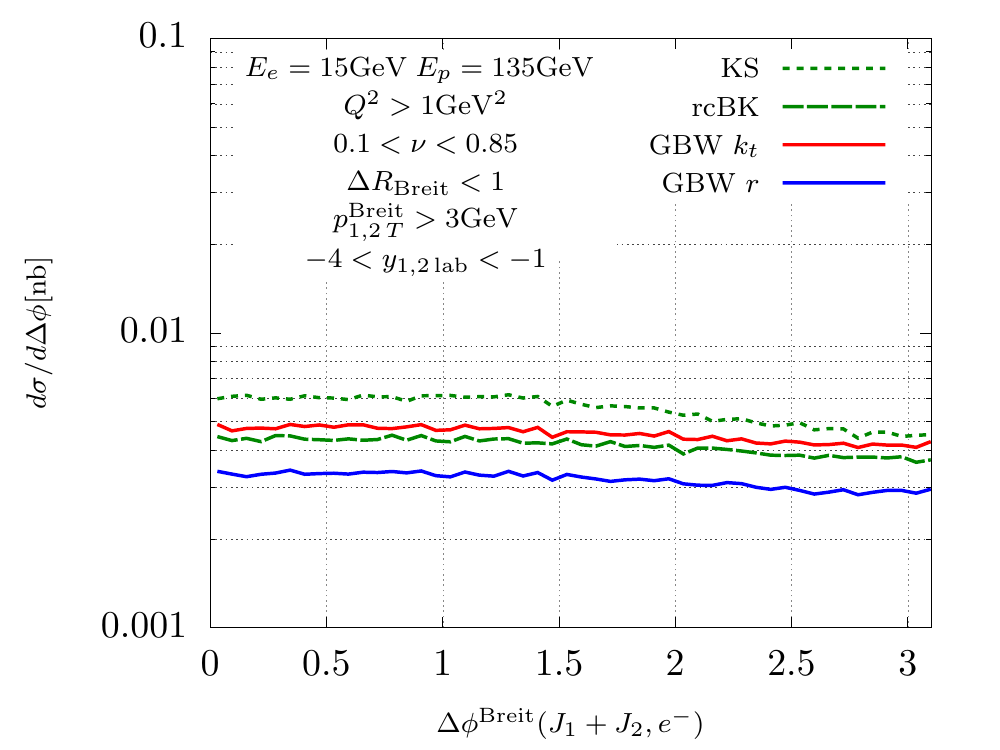} 
	\end{subfigure}
	\begin{subfigure}{0.5\textwidth}
	\includegraphics[width=\textwidth]{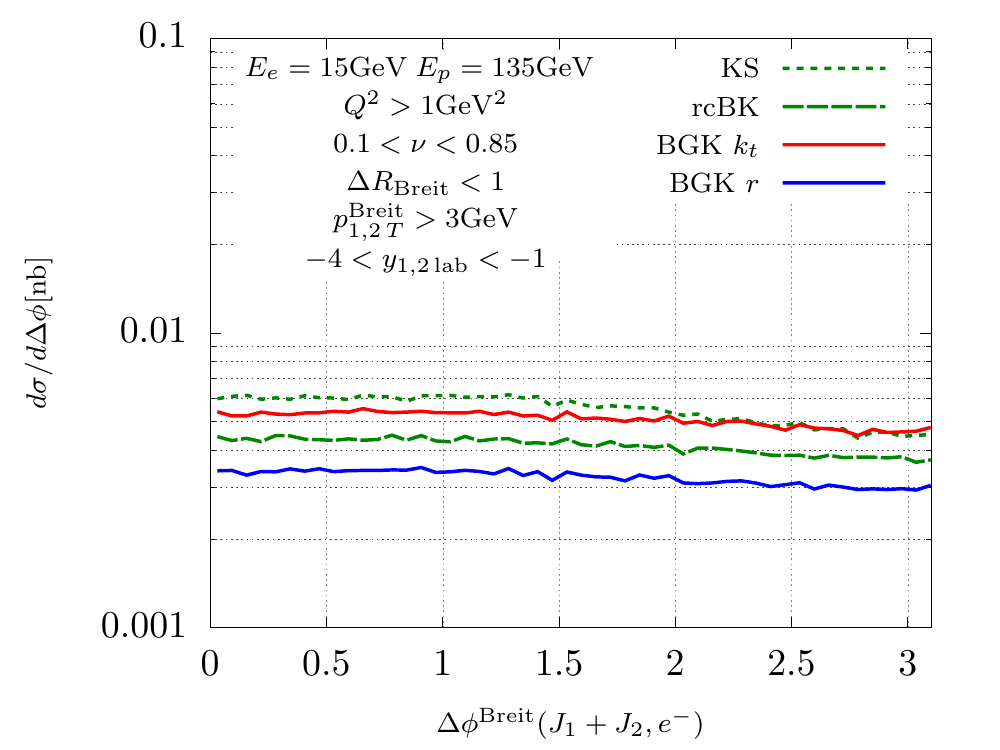} 
	\end{subfigure}
	\begin{subfigure}{0.5\textwidth}
	\includegraphics[width=\textwidth]{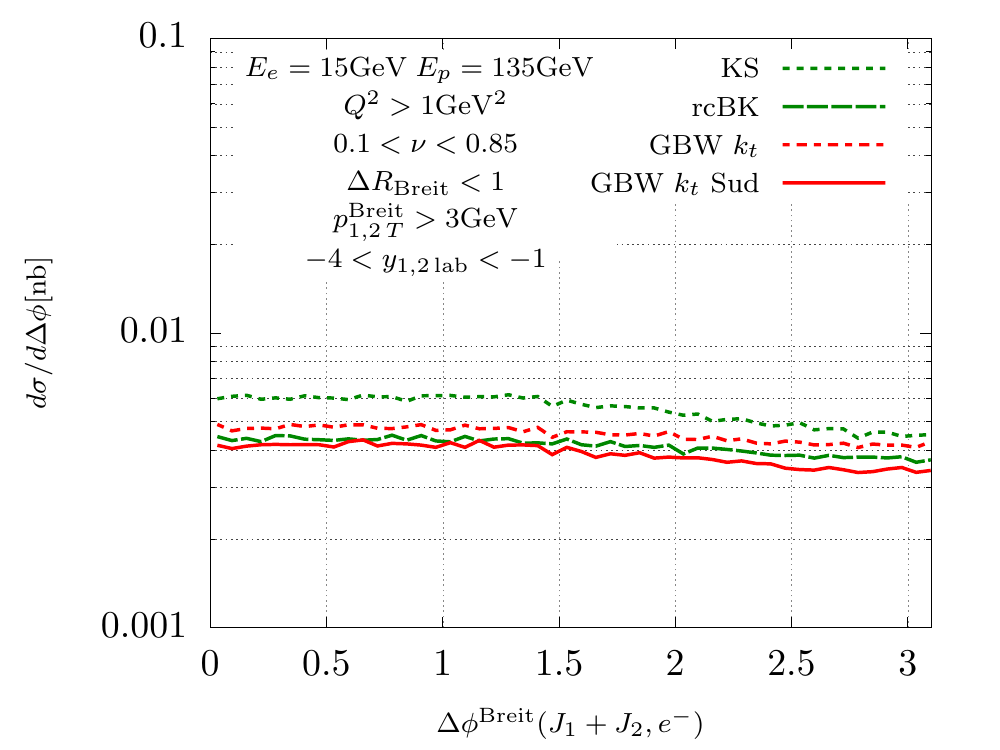}
	\end{subfigure}
	\begin{subfigure}{0.5\textwidth}
	\includegraphics[width=\textwidth]{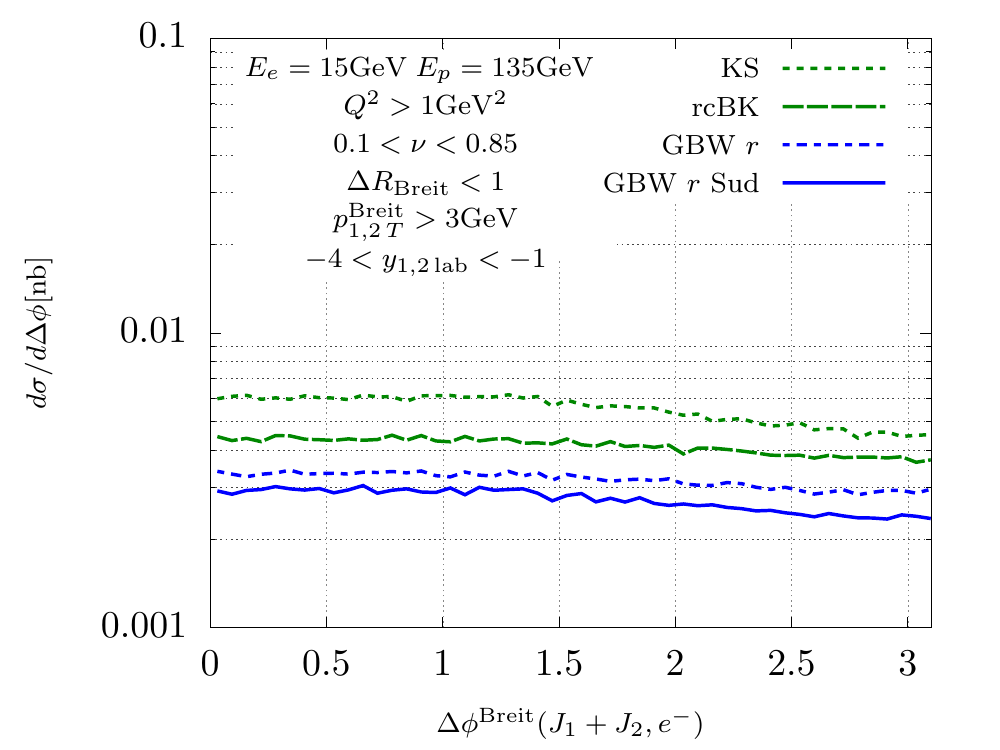}
	\end{subfigure}
	\begin{subfigure}{0.5\textwidth}
	\includegraphics[width=\textwidth]{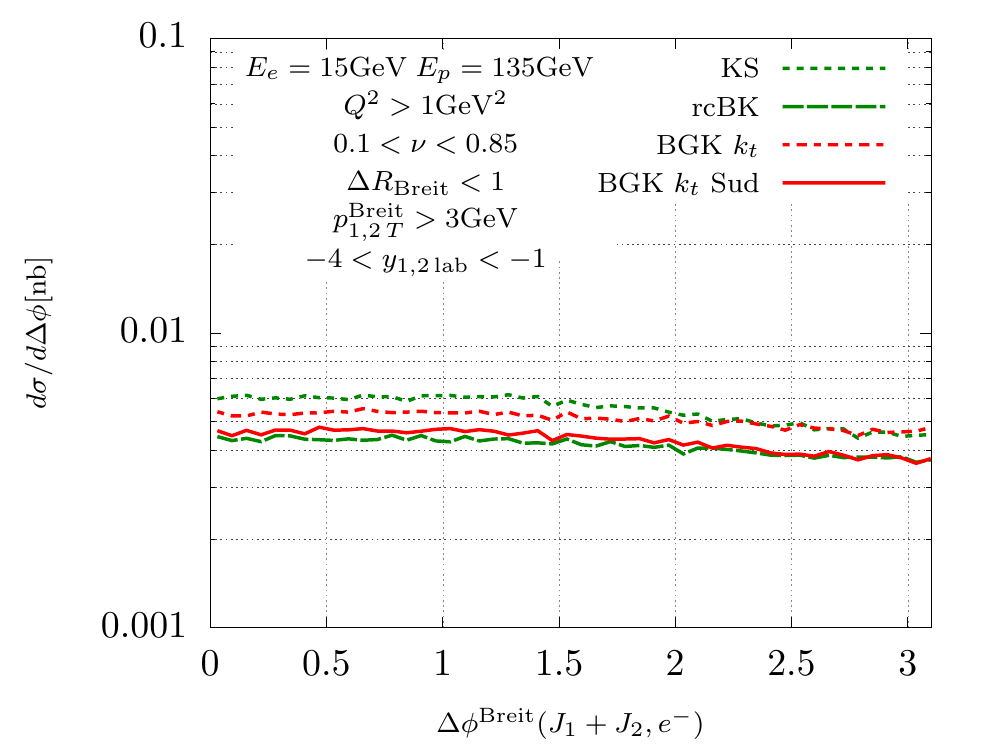}
	\end{subfigure}
	\begin{subfigure}{0.5\textwidth}
	\includegraphics[width=\textwidth]{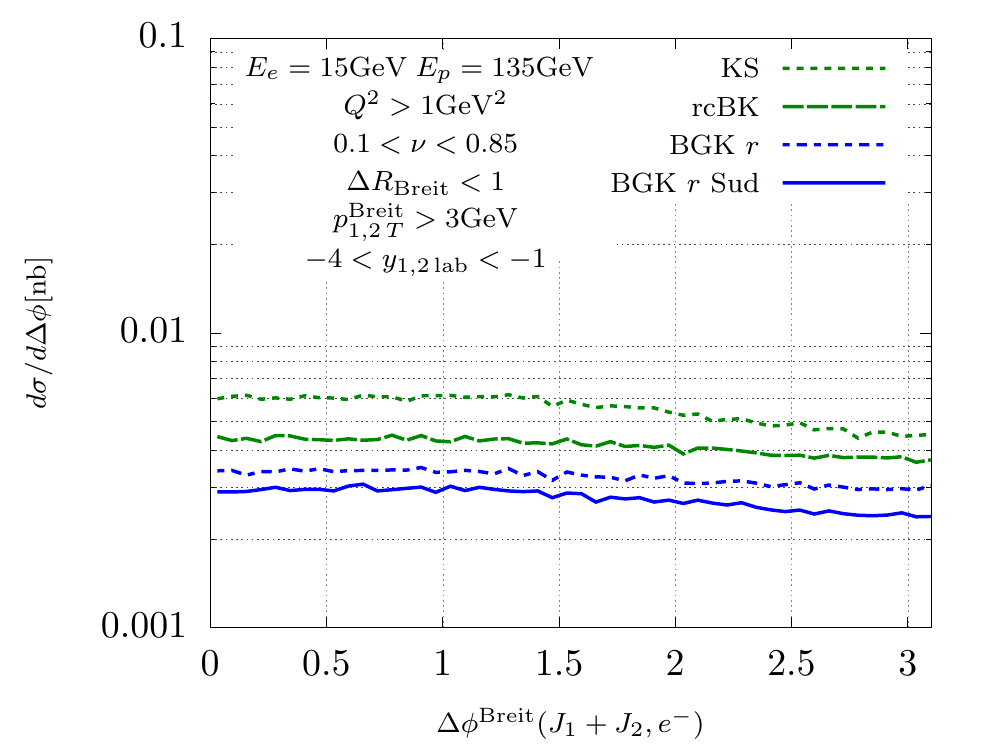}
	\end{subfigure}
	\caption{ Azimuthal correlation of the jets and the scattered electron
	in the Breit frame. Top: Comparison of dipole factorization fit and
	$k_T$-factorization fit. Middle \& Bottom: Effect of the Sudakov form
	factor. The green dotted line is the KS gluon~\cite{vanHameren:2021sqc},
	and the green dashed line is the rcBK
	gluon~\cite{Hentschinski:2022rsa}.}
	\label{fig:je-breit}
\end{figure}

\begin{figure}[p]
	\begin{subfigure}{0.5\textwidth}
		\includegraphics[width=\textwidth]{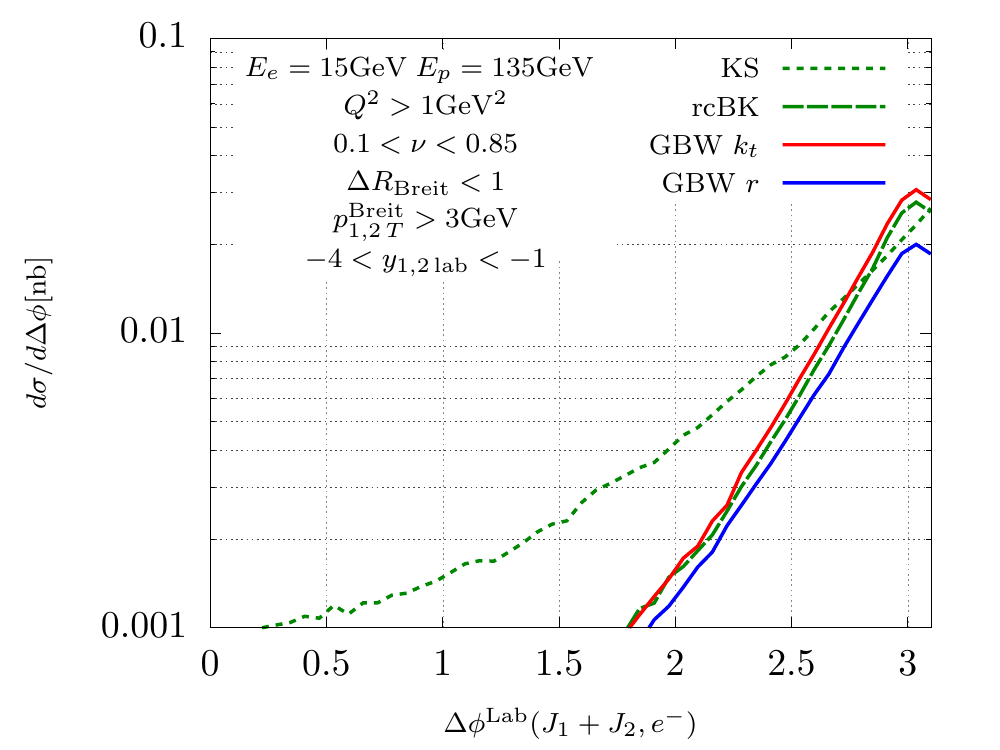} 
	\end{subfigure}
	\begin{subfigure}{0.5\textwidth}
		\includegraphics[width=\textwidth]{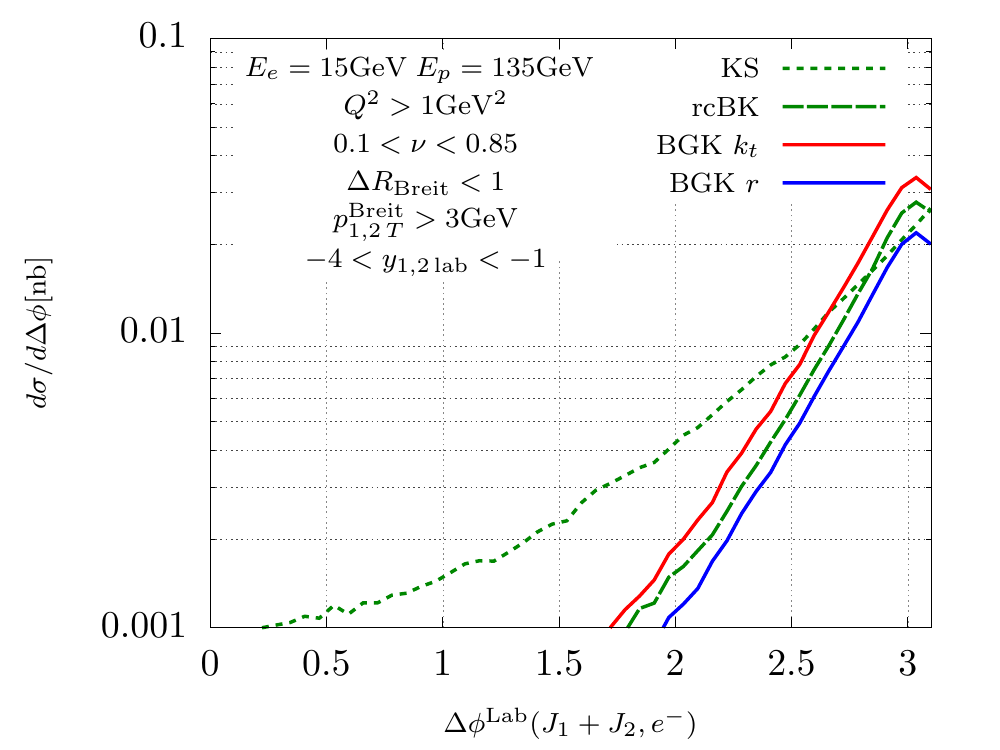} 
	\end{subfigure}

	\begin{subfigure}{0.5\textwidth}
		\includegraphics[width=\textwidth]{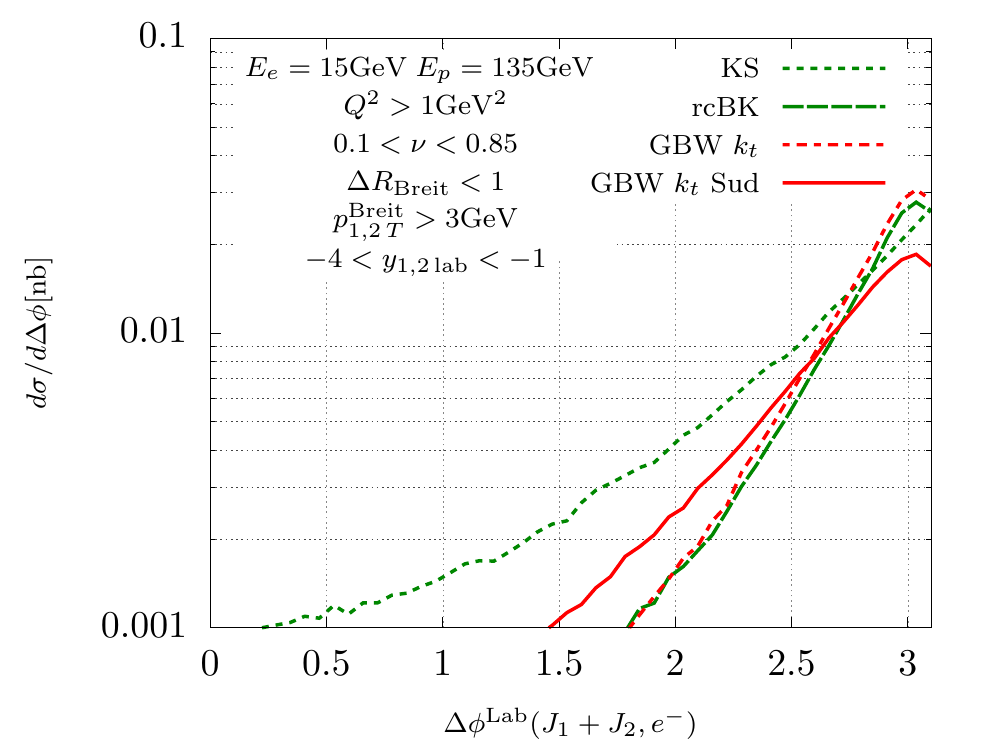}
	\end{subfigure}
	\begin{subfigure}{0.5\textwidth}
		\includegraphics[width=\textwidth]{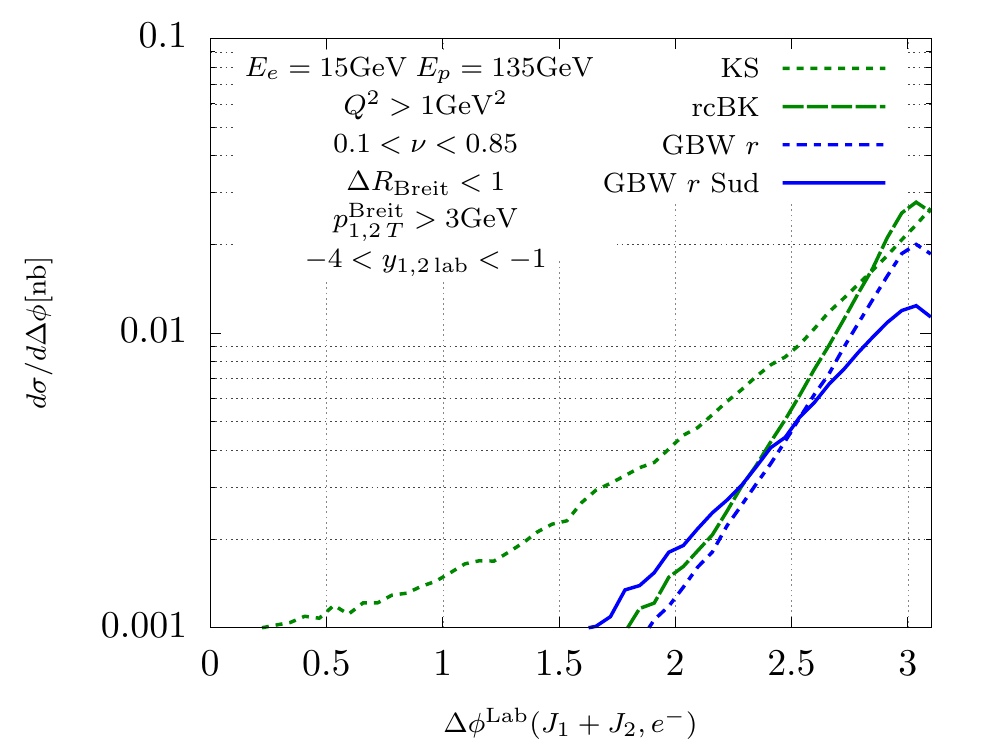}
	\end{subfigure}

	\begin{subfigure}{0.5\textwidth}
		\includegraphics[width=\textwidth]{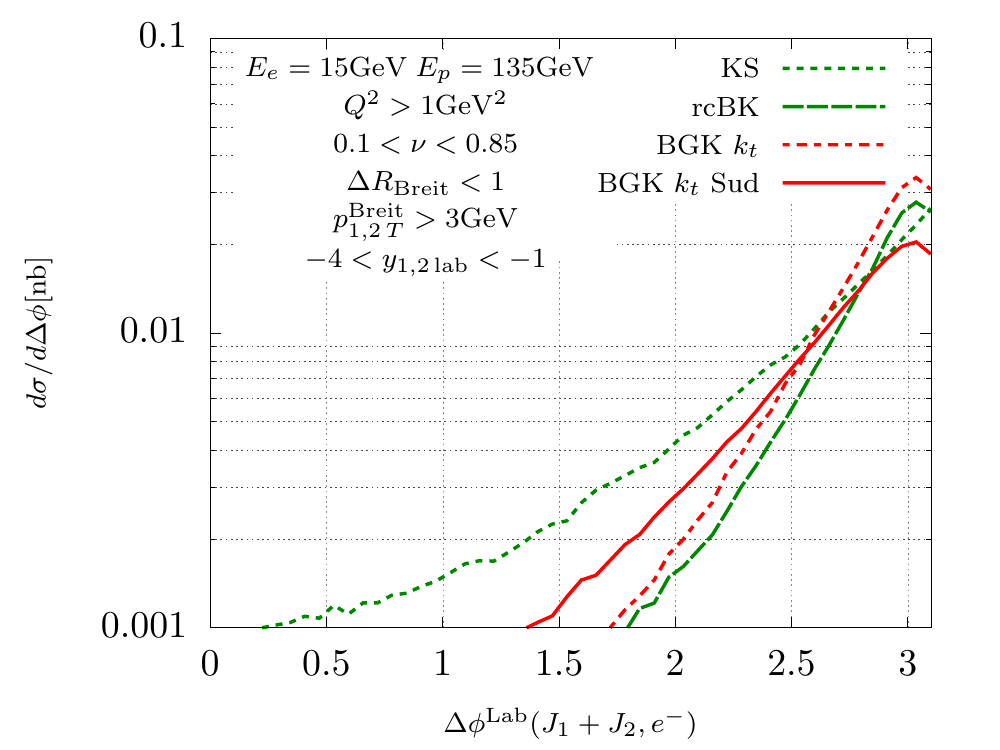}
	\end{subfigure}
	\begin{subfigure}{0.5\textwidth}
		\includegraphics[width=\textwidth]{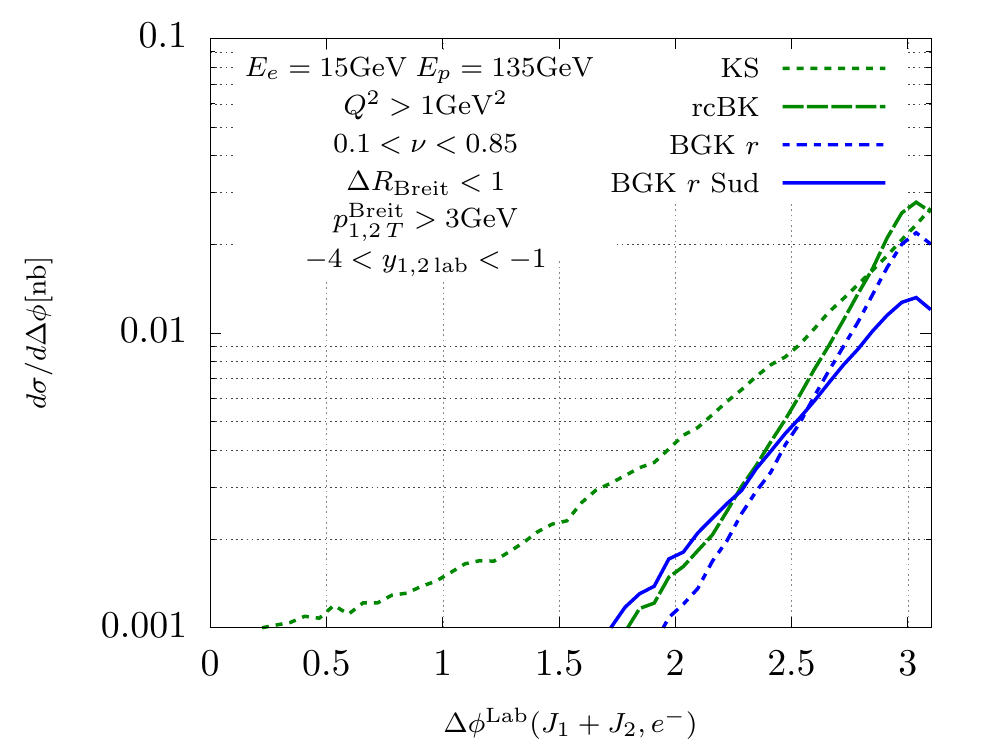}
	\end{subfigure}
	\caption{Azimuthal correlation of the jets and the scattered electron in
	the lab frame. Top: Comparison of dipole factorization fit and
	$k_T$-factorization fit. Middle \& Bottom: Effect of the Sudakov form
	factor. The green dotted line is the KS gluon~\cite{vanHameren:2021sqc},
	and the green dashed line is the rcBK
	gluon~\cite{Hentschinski:2022rsa}.}
        \label{fig:je-lab}
\end{figure}

\begin{figure}[p]
	\begin{subfigure}{0.5\textwidth}
		\includegraphics[width=\textwidth]{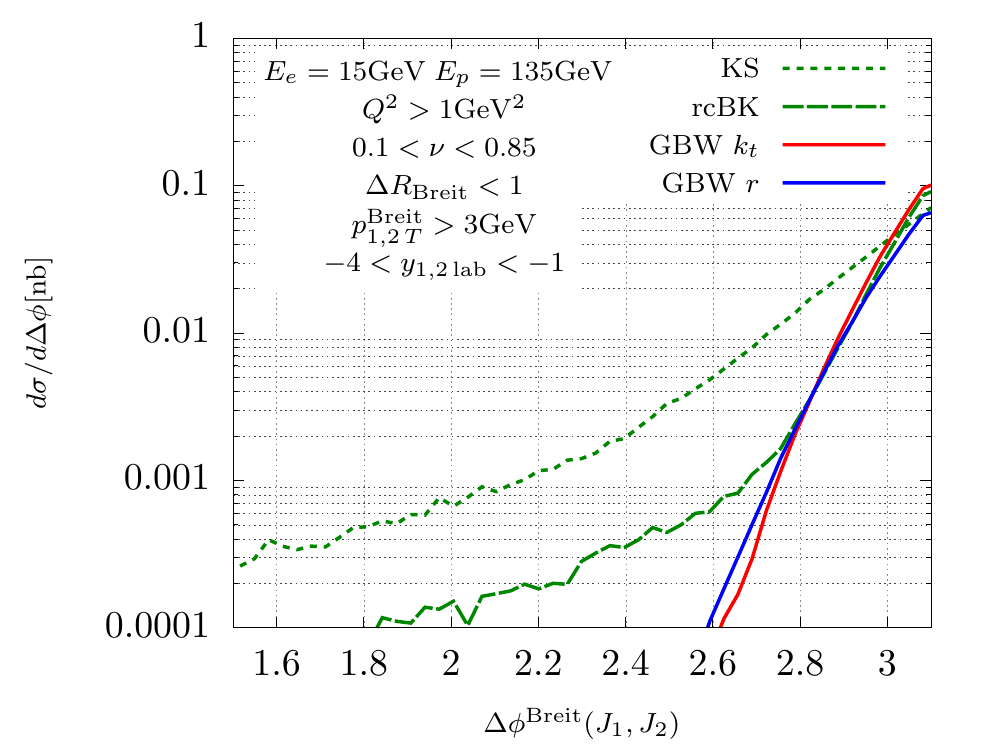} 
	\end{subfigure}
	\begin{subfigure}{0.5\textwidth}
		\includegraphics[width=\textwidth]{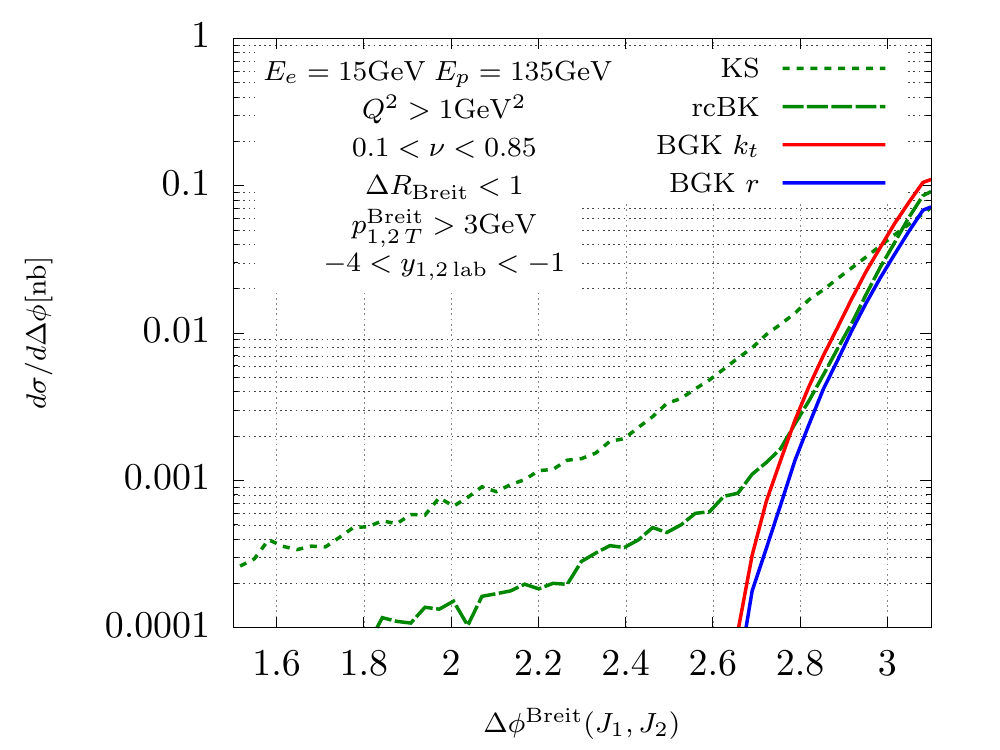} 
	\end{subfigure}

	\begin{subfigure}{0.5\textwidth}
		\includegraphics[width=\textwidth]{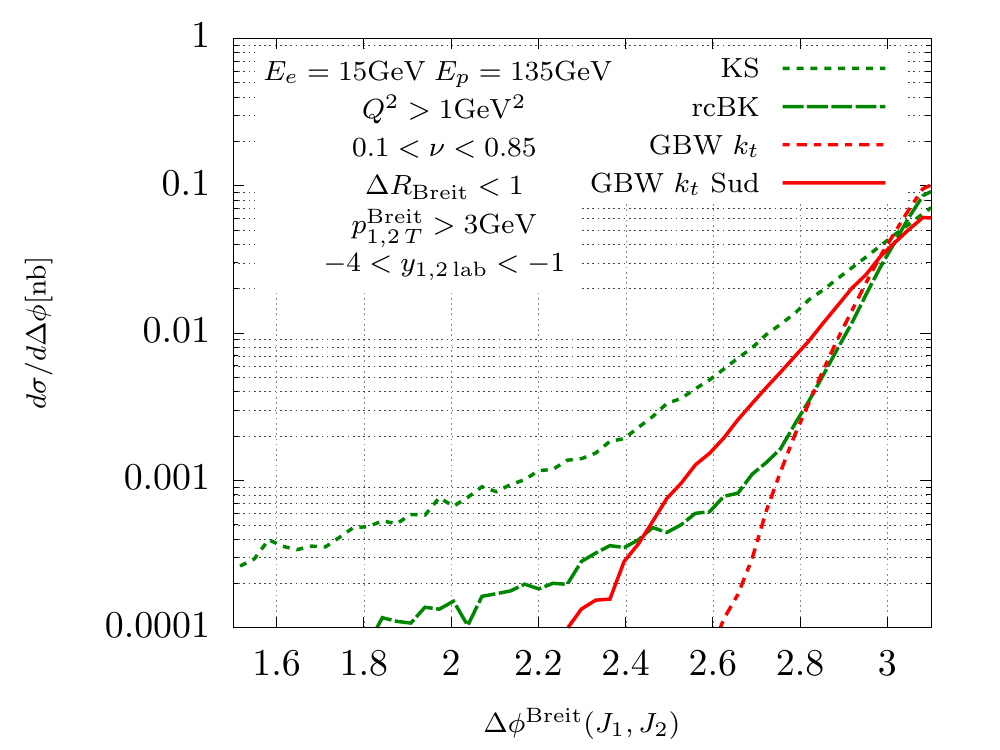}
	\end{subfigure}
	\begin{subfigure}{0.5\textwidth}
		\includegraphics[width=\textwidth]{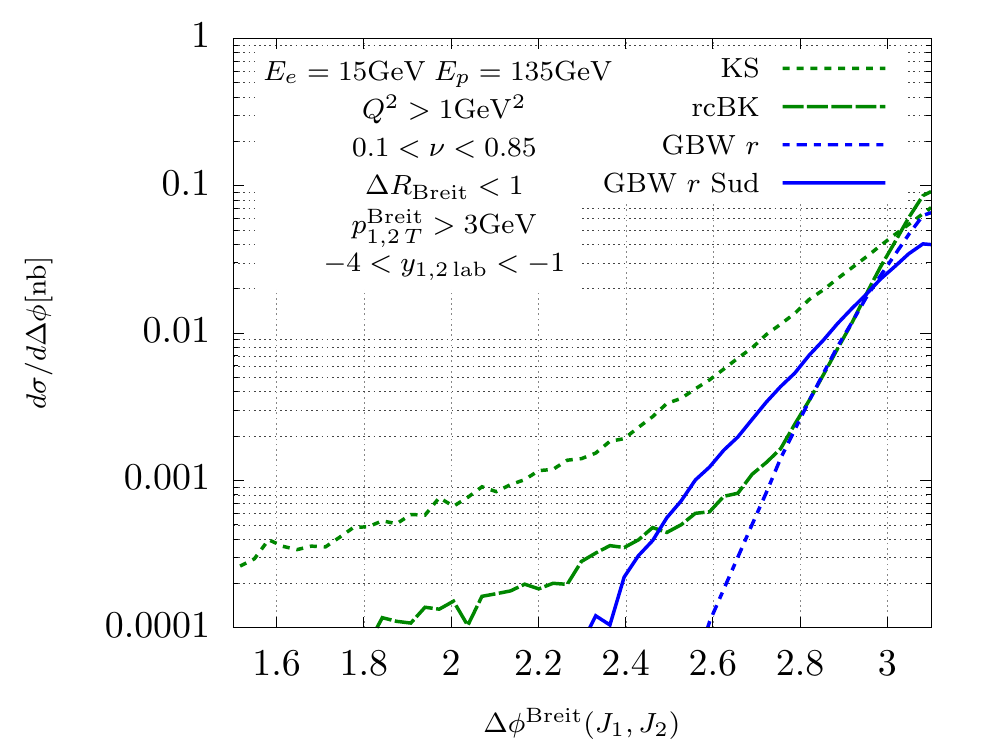}
	\end{subfigure}

	\begin{subfigure}{0.5\textwidth}
		\includegraphics[width=\textwidth]{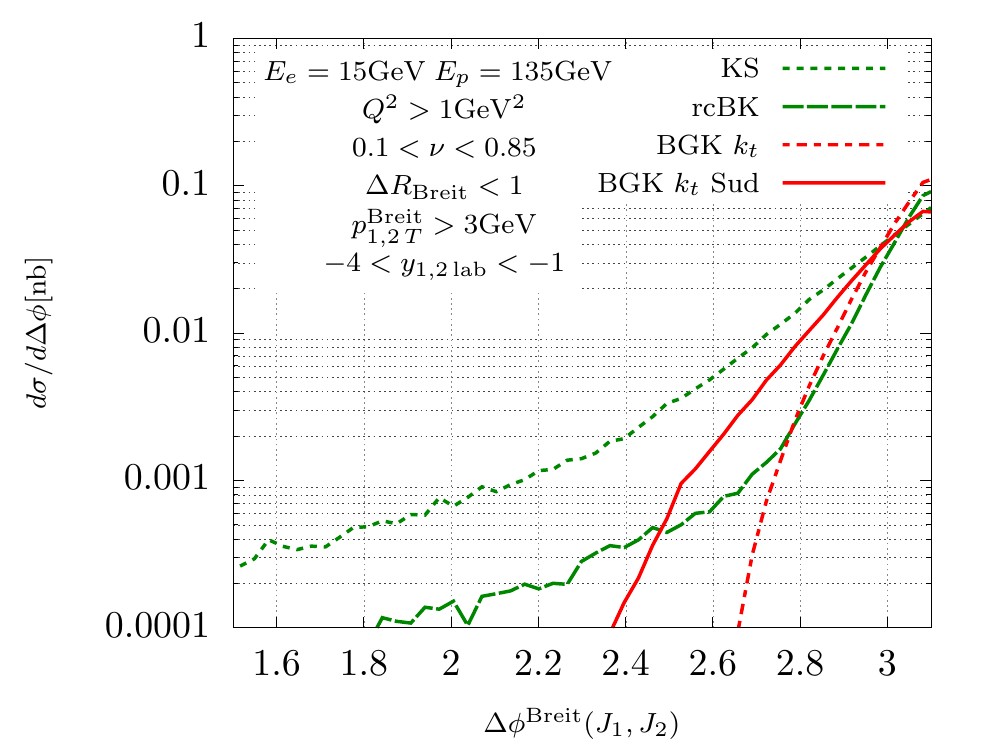}
	\end{subfigure}
	\begin{subfigure}{0.5\textwidth}
		\includegraphics[width=\textwidth]{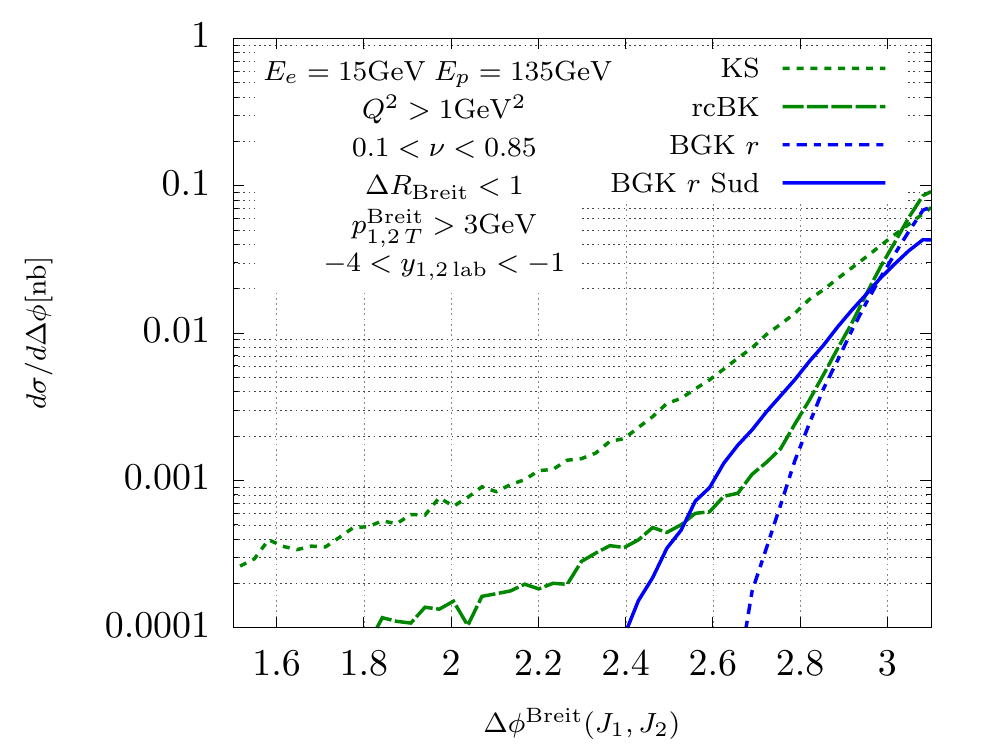}
	\end{subfigure}
	\caption{Azimuthal correlation of the jets in the Breit frame. Top:
	Comparison of dipole factorization fit and $k_T$-factorization fit.
	Middle \& Bottom: Effect of the Sudakov form factor. The green dotted
	line is the KS gluon~\cite{vanHameren:2021sqc}, and the green dashed
	line is the rcBK gluon~\cite{Hentschinski:2022rsa}.}
        \label{fig:jj-breit}
\end{figure}

The $F_2$ structure function is an inclusive object and it is weakly sensitive
to the shape of the gluon density. To probe that shape better, we shall now
apply the dipole cross sections obtained in the previous section to the
jet correlations at the EIC, following closely the method of
Ref.~\cite{vanHameren:2021sqc}.  

We consider dijet production in DIS
\begin{equation}
  e+p\rightarrow e+J_1+J_2+X\,.
\end{equation}
At the leading order, in the small-$x$ limit, this process is dominated by
$q\overline{q}$ jets~\cite{Dominguez:2011wm}.  It is therefore closely related
to the dipole picture we discussed earlier. In the Breit frame, where the photon
momentum is given by $q=(0,0,0,Q)$, at the leading order, the jets momentum
imbalance $p_T\equiv\left|p_{1T}+p_{2T}\right|$ equals the gluon transverse
momentum $k_T$, where $p_{1T}$ and $p_{2T}$ are transverse momenta of the jets.
This makes dijets an interesting process. For the region where $p_T\ll P_T\sim
p_{1T},\,p_{2T}$, one may use power counting to take leading order in $p_T/P_T$,
which leads to the transverse-momentum-dependent~(TMD) factorization.
  
In the large-$N_c$ limit of the TMD factorization, there are two types of gluon
densities, namely the dipole gluon density and the Weizs\"acker-Williams (WW)
gluon
density~\cite{Dominguez:2010xd,Dominguez:2011wm,vanHameren:2016ftb,Xiao:2017ggh}.
It was shown in Ref.~\cite{Dominguez:2011wm}, that the dijet process in DIS can
directly probe the WW gluon, $\fww(x,k^2)$, where the differential cross section
factorizes as
\begin{equation}
	\frac{d\sigma^{\gamma^*p\rightarrow q\overline{q}X}}{d \mathrm{P.S.}}=\fww(x,k_T^2)H_{\gamma^*g^*\rightarrow q\overline{q}},
        \label{eq:TMD}
\end{equation}
with the hard function $H_{\gamma^*g^*\rightarrow q\overline{q}}$ describing
interactions of an off-shell photon with an off-shell gluon producing a
$q\overline{q}$ pair.  $\fww(x,k^2)$, has an interpretation as a number density
of gluons inside a proton, while $\fdp(x,k_T^2)$ does not have such an
interpretation~\cite{Dominguez:2010xd,Dominguez:2011wm,Xiao:2017ggh}.  

We carry out our study in the framework of the {\it Improved
Transverse-Momentum-Dependent} (ITMD)
factorization~\cite{Kotko:2015ura,vanHameren:2016ftb}\footnote {We limit
ourselves to unpolarized contribution as it is the lading one. For the
polarized one,  see Ref.~\cite{Boussarie:2021ybe}}.  This is implemented in
the program KaTie~\cite{vanHameren:2016kkz}, which we use to compute the cross
sections.  The ITMD factorization is a generalization of the TMD factorization,
where the momentum imbalance in TMD is restricted to be
small~\cite{Kotko:2015ura,vanHameren:2016ftb}. That is to say, ITMD resums
$(Q_s/k_T)^n$ and $(k_T/P_T)^n$~\cite{Kotko:2015ura,vanHameren:2016ftb}, thus
extends the region of applicability up to $k_T\sim P_T$. The difference of
Eq.~(\ref{eq:TMD}) from the regular TMD is that the hard function has an
off-shell gluon, $g^*$, thus rendering the $k_T$ dependence in the hard function
as well~\cite{Kotko:2015ura}.

Under the Gaussian approximation and assuming $\theta$-like profile of the
proton, one can
write~\cite{vanHameren:2016ftb,Xiao:2017ggh,Dominguez:2010xd,Dominguez:2011wm}
\begin{equation}
\fww(x,k^2)= \frac{C_F}{2\pi^3\alpha_s}\int^\infty_0\frac{dr}{r}J_0(r k) \sdpa(x,r),
\end{equation} 	
with the adjoint dipole cross section
\begin{equation}
\sdpa(x,r)=\sigma_0\left( 1-\left(1-\frac{\sdp(x,r)}{\sigma_0}\right)^{C_A/C_F}\right).
\label{eq:ww}
\end{equation}

For the region where the TMD factorization is applicable, $Q_s\sim k_T\ll
P_T\sim Q$, one needs to resum the large Sudakov logarithms $\log(k_T/Q)$, as
well as $\log(1/x)$~\cite{Dominguez:2011wm}. It was shown in
Refs.~\cite{Mueller:2012uf,Mueller:2013wwa,Xiao:2017yya} that consistent
resummation of such logarithms is possible owing to the separation of
corresponding regions (see also Ref.~\cite{Nefedov:2021vvy}).  Resummation of
the Sudakov logarithms is achieved by the formula
\begin{equation}
	\fww(x,k^2,\mu^2)= \frac{C_F}{2\pi^3\alpha_s}\int^\infty_0\frac{dr}{r}J_0(r k) e^{-S(r,\mu^2)} \sdpa(x,r),
	\label{eq:ww-sud}
\end{equation}
where, we use the Sudakov form factor~\cite{Mueller:2013wwa,Xiao:2017yya},
\begin{equation}
	S(r,\mu^2)=\frac{\alpha_s N_c}{4\pi}\ln^2\left(\frac{\mu^2r^2}{4e^{-2\gamma_E}}\right),
\end{equation}
in which $\gamma_E$ is the Euler-Mascheroni constant, and we set $\alpha_s=0.2$. 
Following Ref.~\cite{vanHameren:2021sqc}, we study the azimuthal correlations of
jets and the final state electron in DIS , where it was argued that this
observable is sensitive to the soft emissions and the saturation effects. In
this study we focus only on the proton case.  The kinematical cuts suggested in
Ref.~\cite{vanHameren:2021sqc} are
\begin{align*}
	E_e&=15\GeV& E_p&=135\GeV& Q^2&>1\GeV^2\\
	0.1<&\nu<0.85&\Delta R_{\mathrm{Breit}}&<1&p^{\mathrm{Breit}}_{1,2\,T}&>3\GeV\\
	&&-4<&y_{1,2\,\mathrm{lab}}<-1.&&
\end{align*}

Grids of the Weizs\"acker-Williams gluon density were produced by evaluating
Eqs.~(\ref{eq:ww}) and~(\ref{eq:ww-sud}).  The gluon density at $x=10^{-3}$ is
plotted in Fig.~\ref{fig:ww} with the hard-scale-independent Kutak-Sapeta (KS)
gluon~\cite{Kutak:2012rf,vanHameren:2021sqc,Abdulov:2021ivr} and
the running-coupling BK~(rcBK) gluon
density~\cite{Balitsky:2006wa,Albacete:2010sy,Hentschinski:2022rsa}. Both of
these gluon densities are solutions of evolution equations and treat better
perturbative tail at large $k_T$. Furthermore, the KS gluon takes into account
resummed corrections of higher orders, {\it i.e.} kinematical constraint and
nonsingular (at low $z$) elements of DGLAP  splitting
functions~\cite{Kwiecinski:1997ee}. 

Clearly, as shown in Fig.~\ref{fig:ww}, the GBW and BGK models fall much more
quickly than the KS and rcBK gluon densities. In general, expected behaviour in
the large-$k_T$ region is $\sim
k_T^{-2}$~\cite{Dominguez:2010xd,Dominguez:2011wm}, while
$\sigma_{\mathrm{GBW}}$ behaves like $\sim e^{-k_T^2}$.  As in
Ref.~\cite{vanHameren:2021sqc}, the Sudakov factor enhances in the small-$k_T$
region and suppresses in the large-$k_T$ region. In other words, it broadens the
distribution. In comparison to the result of Ref.~\cite{vanHameren:2021sqc}, the
effect of broadening by the Sudakov factor is significantly more pronounced in
the case of the GBW and BGK models. The hard-scale-dependent GBW and BGK models,
as a consequence, become closer to the KS and rcBK gluons, \textit{cf.} Fig.~2
of Ref.~\cite{vanHameren:2021sqc}.

Figs.~\ref{fig:je-breit} and~\ref{fig:je-lab} show electron-jets azimuthal
correlation in the Breit and in the Lab frame, respectively.  In the top row of
Fig.~\ref{fig:je-breit}, we see, for both the GBW and BGK models, better
agreements of results with the KS and rcBK for the new $k_T$-factorization fits.
However the overall normalization of the gluon density depends on the coupling
$\alpha_s$, which we assumed to be~0.2.  Nevertheless, it shows clearly the
effect of the parameter $\sigma_0$. In the middle and the bottom row, it shows
the effects of the Sudakov form factor, which qualitatively agrees with that of
Ref.~\cite{vanHameren:2021sqc}, by lowering the cross section.  

Fig.~\ref{fig:je-lab} shows the electron-jets correlation in the Lab frame.
Here, the difference between KS and GBW and BGK is more prominent, while rcBK
shows similar pattern to the GBW and BGK models and therefore we can attribute
the effects to importance of higher order corrections, as are accounted for in
KS gluon. The effects of the Sudakov factor are similar to those in
Ref.~\cite{vanHameren:2021sqc} at relatively high $\Delta \phi$, while at
smaller $\Delta \phi$, the effect is reversed ({\it i.e.} the cross sections
were slightly lowered in Ref.~\cite{vanHameren:2021sqc}, while here, they are
significantly increased).

Finally, Fig.~\ref{fig:jj-breit} shows the jet-jet correlations in the Breit
frame. Again, the GBW and BGK models exhibit considerable deviation from the KS
gluon. The difference from the previous plot is the disagreement of the GBW/BGK
models and the rcBK in the small-$\Delta\phi$ region. Similarly to the previous
plots, the Sudakov factor affects the models somewhat differently from the KS
gluon in Ref.~\cite{vanHameren:2021sqc}.  The effect enhances the cross section
considerably in the small-$\Delta\phi$ region, making it closer to KS gluon
result.  

The results shown in Figs.~\ref{fig:je-lab} and~\ref{fig:jj-breit} are natural,
as back-to-back configuration in the respective observable corresponds to the
small-$k_T$ region of gluon densities and, as it can be seen clearly in
Fig.~\ref{fig:ww}, the GBW and BGK gluons do not fare well in the large-$k_T$
region.  That is to say that the enhancement in the small-$\Delta\phi$ region is
a direct consequence of the broadening by the Sudakov factor. 

\section{Summary}

We have fitted the GBW and BGK saturation models to HERA data~\cite{Abt:2017nkc}
using the $k_T$-factorization for the structure function $F_2$.  The main
difference between the $k_T$- and the dipole factorizations is an argument of
the gluon, $x/z$, appearing in the former and being replaced by
$x\,(1+4m_f^2/Q^2)$ in the latter. 
In fact, the massive light quarks used in Ref.~\cite{Golec-Biernat:1998zce}
partially simulate the factor $1/z$, and our fit result indicates such effect,
as expected. 
 
We found that the dipole factorization can reproduce the result of the
$k_T$-factorization formula quite well. The only major difference is in the
normalization parameter $\sigma_0$, which increases significantly for the
$k_T$-factorization case. We argued that this change is a direct consequences of
the kinematic approximation used in the dipole factorization formula. We have
also observed that the explicit inclusion of the running coupling in the GBW
model has significant effect on fit quality, particularly in the large-$Q^2$
region, where the GBW model performs poorly. 

We have applied the new results from our fits for predictions of the dijet
process in DIS at EIC. Additionally, effects of the Sudakov form factor were
investigated for that process. Results of the electron-jets correlation in the
Breit frame agree qualitatively with those of Ref.~\cite{vanHameren:2021sqc}.
Other results, namely the electron-jets in the Lab frame and the jet-jet
correlation in the Breit frame, show considerable effects of the Sudakov form
factor, which broadens the gluon density. 

\section*{Acknowledgment}
We are grateful to Krzysztof Golec-Biernat, Piotr Kotko and Andreas van Hameren
for useful discussions. The project is partially supported by the European
Union’s Horizon 2020 research and innovation program under grant agreement No.
824093. TG and SS are partially supported by the Polish National Science Centre
grant no. 2017/27/B/ST2/02004.

\appendix
\section{$I_i$ in Eq.~(\ref{eq:angle-integrated})}

The functions $I_1,\, I_2,\,  I_3,\,  I_4$ used in
Eq.~(\ref{eq:angle-integrated}) are defined  as~\cite{Kimber:2001uaa}
\begin{equation}
	\begin{split}
		\frac{I_1}{2\pi}=\frac{N_1N_2+N_3^2}{\left( N^2_1+2N_1N_2+N_3^2\right)^{3/2}}, &\hspace{1cm}
		\frac{I_2}{2\pi}=\frac{N_3-(1-2\beta)N_1}{(N_1+N_4)\sqrt{ N^2_1+2N_1N_2+N_3^2}},\\
		\frac{I_3}{2\pi}=\frac{N_1+N_2}{\left( N^2_1+2N_1N_2+N_3^2\right)^{3/2}},&\hspace{1cm}
		\frac{I_4}{2\pi}=\frac{2(1-\beta)}{(N_1+N_4)\sqrt{ N^2_1+2N_1N_2+N_3^2}},
	\end{split}
\end{equation}
for
\begin{equation}
	\begin{split}
		N_1\equiv\beta(1-\beta)Q^2+m_f^2, &\hspace{2cm}
		N_2\equiv{\kappa'}_t^2+(1-\beta)^2k_T^2,\\
		N_3\equiv{\kappa'}_t^2-(1-\beta)^2k_T^2, &\hspace{2cm}
		N_4\equiv{\kappa'}_t^2+\beta(1-\beta)k_T^2.
	\end{split}
\end{equation}


\bibliographystyle{unsrt}


\end{document}